\documentclass[preprint,prd,aps,showpacs,showkeys,nofootinbib]{revtex4}
\usepackage{graphicx}
\begin{document}
\title{Gravitational waves from the phase transition in the B-LSSM}

\author{Xing-Xing Dong$^{1,2}$\footnote{dxx$\_$0304@163.com},Tai-Fu Feng$^{1,2,3}$\footnote{fengtf@hbu.edu.cn},Hai-Bin Zhang$^{1,2}$\footnote{hbzhang@hbu.edu.cn},Shu-Min Zhao$^{1,2}$\footnote{zhaosm@hbu.edu.cn},Jin-Lei Yang$^{1,2,4}$\footnote{yangjinlei@itp.ac.cn}}

\affiliation{$^1$ Department of Physics, Hebei University, Baoding 071002, China\\
$^2$ Key Laboratory of High-precision Computation and Application of Quantum Field Theory of Hebei Province, China\\
$^3$ Department of Physics, Chongqing University, Chongqing 401331, China\\
$^4$ University of Chinese Academy of Sciences, Beijing 100049, China}

\begin{abstract}
Based on the gauge symmetry group $SU(3)_C\otimes{SU(2)_L}\otimes{U(1)_Y}\otimes{U(1)_{B-L}}$, the minimal supersymmetric extension of the SM with local B-L gauge symmetry(B-LSSM) has been introduced. In this model, we study the Higgs masses with the one-loop zero temperature effective potential corrections. Besides, the finite temperature effective potentials connected with two $U(1)_{B-L}$ Higgs singlets are deduced specifically. Then we can obtain the gravitational wave spectrums generated from the strong first-order phase transition. In the B-LSSM, the gravitational wave signals can be as strong as $h^2\Omega_{GW}\sim10^{-11}$, which may be detectable in the future experiments.
\end{abstract}


\maketitle
\section{introduction}
Based on the gauge group $SU(3)_C\otimes{SU(2)_L}\otimes{U(1)_Y}$, the standard model(SM) has been successfully established. Nevertheless, the SM possesses some limitation and additional $U(1)_{B-L}$ gauge interaction will be a promising extension to the SM. Therefore, the minimal supersymmetric extension of the SM(MSSM)\cite{MSSM1,MSSM2,MSSM3,MSSM4} with local B-L gauge symmetry (B-LSSM) is introduced, where $B$ represents baryon number and $L$ stands for lepton number\cite{B-LSSM1,B-LSSM2,B-LSSM3,B-LSSM4,B-LSSM5}. In this model, the invariance under $U(1)_{B-L}$ gauge group imposes the R-parity conservation to avoid proton decay\cite{B-L R Parity}. Besides, the right-handed neutrino superfields have been imported in the B-LSSM to obtain the tiny neutrino masses through type-I seesaw mechanism. Thus, B-LSSM provides an elegant solution to the existence of the light left-handed neutrino. Then the possibility of baryogenesis via leptogenesis will be well explained. Furthermore, due to the introduction of the additional singlet Higgs states and right-handed (s)neutrinos, B-LSSM alleviates the hierarchy problem through additional parameter space released from the LEP, Tevatron and LHC constraints\cite{B-L hierarchy1,B-L hierarchy2}. Other than this, the model can also provide much more dark matter candidates\cite{B-LDM1,B-LDM2,B-LDM3,B-LDM4} than the MSSM.

The gravitational wave(GW) signals have been detected at the Laser Interferometer Gravitational Wave Observer (LIGO)\cite{LIGO1,LIGO2}, which urges physicists to explore the various universe mysteries. The sources of GW signals can be sensitive to arise from the first-order phase transition(PT) in the early universe\cite{GWsource1,GWsource2,GWsource3,GWsource4,GWsource5}. The electroweak symmetric broken phase is produced in the form of bubbles. When the universe cools down to the nucleation temperature, bubbles of broken phase may nucleate in the background of symmetric phase. Bubbles expand, collide, merge and finally fill the whole universe to finish the first-order PT. Then, the stochastic GW signals can be generated through the bubble collisions, sound waves after the bubble collision and magnetohydrodynamic turbulent of the surrounding plasma\cite{GWsource4,GWgeneration2,GWgeneration3}.

Unfortunately, the electroweak PT in the SM turns out
to be too weak to obtain GW signals\cite{GWSM1,GWSM2,GWSM3}. In order to realize strong first-order PT and finally obtain the GW spectrums, physicists have studied various extensions of the SM, such as the model with a dimension-six
operator\cite{GWsixoperator1,GWsixoperator2,GWsixoperator3}, NMSSM\cite{GWNMSSM1,GWNMSSM2,GWNMSSM3} and other studies\cite{GWothermodel1,GWothermodel2,GWothermodel3}. In this paper, we hope to investigate the GW spectrums produced from the first-order PT of the B-LSSM. We find that the strength of GW signals will be up to $h^2\Omega_{GW}\sim10^{-11}$, which can be detected by the future experiments, such as the LISA with N2A5M5L6 design configurations, the Big Bang Observer(BBO), DECi-hertz Interferometer Observatory(DECIGO) and Ultimate-DECIGO\cite{GWexp1,GWexp2}.

In this paper, after discussing the Higgs mass, we mainly study the GW spectrums generated from the strong first-order PT of the B-LSSM. We first introduce characteristics of the B-LSSM, then we discuss the Higgs masses corrected by the one-loop zero temperature effective potential in section II. In section III, we compute the concrete $(\phi_{\eta},\phi_{\bar{\eta}})$-dependent finite temperature effective potential. Meanwhile, we derive the GW generation by cosmological first-order PT in section IV. The GW spectrums are comprised by the bubble collision, sound wave and turbulence contributions. The numerical results of the GW spectrums that depend on the model parameters will be further illustrated in section V. Last but not least, we summarize the conclusion in section VI. The anomaly free of B-LSSM will be demonstrated in Appendix A.
\section{the B-LSSM and the Higgs mass}
\subsection{The B-LSSM}
In 1967, Sakharov proposed three necessary conditions for dynamics to produce asymmetry between matter and anti-matter in the universe\cite{asymmetrycondition}: (1) baryon number(B) non-conservation; (2) charge conjugation(C) transformation and charge conjugate-parity(CP) joint transformation non-conservation; (3) The system deviation from thermal equilibrium. Through a lot of research, people realize that the SM can satisfy Sakharov's three conditions. However, the CP violation given by the CKM matrix in the SM is not sufficient to account for the observed baryon-photon number density ratio. In addition, in order to produce strong first-order electroweak phase transition, the mass of Higgs particle in the SM must be less than 45 GeV. In 2012, the mass of physical Higgs particle was found to be around 125 GeV by the European Large Hadron Collider (LHC)\cite{h0ATLAS,h0CMS}, which directly denied the possibility of realizing the electroweak baryon number production mechanism (baryogenesis) in the SM. Therefore, the current SM can not explain the matter and anti-matter asymmetries in the universe.

In the B-LSSM, the baryon number and lepton number are broken by Sphaleron process respectively, but the difference between the two is conserved. This links the change of the baryon number with the change of the lepton number, and the asymmetry of the baryon number can be converted from the asymmetry of the lepton number through the sphaleron process. In general, the mechanism of lepton number asymmetry (leptogenesis) requires the lepton number violating process, the C and CP destruction of the lepton part and the realization of the non-equilibrium state. These conditions can be achieved in the general models within the mass neutrino. The B-LSSM introduces the right-handed neutrinos. Besides, neutrinos are Majorana type and gain mass through the type-I seesaw mechanism, which breaks the lepton number symmetry. In B-LSSM, the C or CP symmetry is broken. In addition, the decoupling of heavy right-handed neutrinos provides non-equilibrium conditions. Sphhaleron process only acts directly on left-handed fermions, and can partially convert the lepton number of left-handed leptons to the baryon number, thus explains the baryon number asymmetry. Therefore, the possibility of baryogenesis via leptogenesis can be realized within the B-LSSM.

In the B-LSSM, the local gauge group is defined as $SU(3)_C\otimes{SU(2)_L}\otimes{U(1)_Y}\otimes{U(1)_{B-L}}$. The specification of the quantum numbers of fields in the B-LSSM will be discussed in the TABLE \ref{B-Lquantum numbers}. Besides, B-LSSM is assumed the gauged one, and we have proved carefully that this model is anomaly free. The concrete demonstrations will be given out in the Appendix A. B-LSSM introduces $U(1)_{B-L}$ gauge field, right-handed neutrinos $\nu_i^c$ and their superpartners. The masses of right-handed neutrinos will be constructed by $Y_{x,ij}\hat{\nu}_i^c\hat{\eta}\hat{\nu}_j^c$. Then we can obtain the tiny neutrinos masses through the type-I seesaw mechanism after the right-handed neutrinos and left-handed neutrinos mixing together. The B-LSSM superpotential is deduced as
\begin{table}[t]
\footnotesize
\begin{tabular}{|c|c|c|c|c|c|}
\hline
SF & Spin 0 & Spin \(\frac{1}{2}\) & Generations & \((U(1)_Y\otimes\, \text{SU}(2)\otimes\, \text{SU}(3)\otimes\, U(1)_{B-L})\) \\
\hline
\(\hat{H}_d\) & \(H_d\) & \(\tilde{H}_d\) & 1 & \((-\frac{1}{2},{\bf 2},{\bf 1},0) \) \\
\(\hat{H}_u\) & \(H_u\) & \(\tilde{H}_u\) & 1 & \((\frac{1}{2},{\bf 2},{\bf 1},0) \) \\
\(\hat{Q}\) & \(\tilde{Q}\) & \(Q\) & 3 & \((\frac{1}{6},{\bf 2},{\bf 3},\frac{1}{6}) \) \\
\(\hat{L}\) & \(\tilde{L}\) & \(L\) & 3 & \((-\frac{1}{2},{\bf 2},{\bf 1},-\frac{1}{2}) \) \\
\(\hat{D}\) & \(\tilde{D}_R^{*}\) & \(D_R^*\) & 3 & \((\frac{1}{3},{\bf 1},{\bf \overline{3}},-\frac{1}{6}) \) \\
\(\hat{U}\) & \(\tilde{U}_R^{*}\) & \(U_R^*\) & 3 & \((-\frac{2}{3},{\bf 1},{\bf \overline{3}},-\frac{1}{6}) \) \\
\(\hat{R}\) & \(\tilde{R}^*\) & \(R^*\) & 3 & \((1,{\bf 1},{\bf 1},\frac{1}{2}) \) \\
\(\hat{\nu}\) & \(\tilde{\nu}_R^*\) & \(\nu_R^*\) & 3 & \((0,{\bf 1},{\bf 1},\frac{1}{2}) \) \\
\(\hat{\eta}\) & \(\eta\) & \(\tilde{\eta}\) & 1 & \((0,{\bf 1},{\bf 1},-1) \) \\
\(\hat{\bar{\eta}}\) & \(\bar{\eta}\) & \(\tilde{\bar{\eta}}\) & 1 & \((0,{\bf 1},{\bf 1},1) \) \\
\hline
\end{tabular}
\caption{ \label{B-Lquantum numbers}  The quantum numbers of fields in the B-LSSM.}
\end{table}
\begin{eqnarray}
W_{B-L} ={\cal W}_{MSSM}- {\mu_{\eta}} \hat{\eta}\hat{\bar{\eta}}+Y_{x,ij}\hat{\nu}_i^c\hat{\eta}\hat{\nu}_j^c+Y_{\nu,ij}\hat{L}_i\hat{H}_u\hat{\nu}_j^c,
\end{eqnarray}
where ${\cal W}_{MSSM}$ is the superpotential of MSSM. $i, j$ represent the generation indices, $Y_{x,ij}$ and $Y_{\nu,ij}$ correspond to the Yukawa coupling coefficients. $\mu_{\eta}$, considered as a parameter with mass dimension, is the supersymmetric mass between $U(1)_{B-L}$ Higgs singlets $\eta$ and $\bar{\eta}$.

When the Higgs fields acquire nonzero vacuum expectation values (VEVs), the gauge group
$SU(3)_C\bigotimes SU(2)_L\bigotimes U(1)_Y\bigotimes U(1)_{B-L}$ is broken down into electromagnetic symmetry $U(1)_{em}$:
\begin{eqnarray}
&&H_d^0 = \frac{1}{\sqrt{2}}( \phi_d + v_d  + i \sigma_d),\;\;
H_u^0 = \frac{1}{\sqrt{2}} (\phi_u  +  v_u  + i \sigma_u),
\nonumber\\&&\eta = \frac{1}{\sqrt{2}}( \phi_{\eta} + v_{\eta}  + i \sigma_{\eta} ),\;\;\;\;\;
\bar{\eta} = \frac{1}{\sqrt{2}} (\phi_{\bar{\eta}}  +  v_{\bar{\eta}}  + i \sigma_{\bar{\eta}}),
\end{eqnarray}
where $\phi_d,\;\phi_u,\;\phi_{\eta},\;\phi_{\bar{\eta}}$ represent the CP-even Higgs components and $\sigma_d,\;\sigma_u,\;\sigma_{\eta},\;\sigma_{\bar{\eta}}$ correspond to the CP-odd Higgs components. The VEVs of the $U(1)_{B-L}$ Higgs singlets $\hat{\eta}$ and $\hat{\bar{\eta}}$ satisfy $u=\sqrt{v_{\eta}^2+v_{\bar{\eta}}^2}$. While the VEVs of the Higgs doublets $\hat{H}_d$ and $\hat{H}_u$ are $v_d$ and $v_u$, which satisfy $v=\sqrt{v_d^2+v_u^2}$. We take $tb'=\frac{v_{\bar{\eta}}}{v_\eta}$ by analogy to the definition $tb=\frac{v_u}{v_d}$ in the MSSM.
\subsection{Higgs mass in the B-LSSM}
First, in the base $(\phi_d,\phi_u,\phi_\eta,\phi_{\bar{\eta}})$, the tree level mass squared matrix for Higgs boson $M_h^2$ is deduced as:
\begin{eqnarray}
M_h^2=u^2\times\hspace{-0.1cm}
\left(\begin{array}{*{20}{c}}
{\frac{1}{4}\frac{g^2 x^2}{1+tb^2}\hspace{-0.1cm}+\hspace{-0.1cm}n^2tb}&{-\frac{1}{4}g^2\frac{x^2tb}{1+tb^2}}\hspace{-0.1cm}-\hspace{-0.1cm}n^2&
{\frac{1}{2}g_{_B}g_{_{YB}}\frac{x}{T}}&
{-\frac{1}{2}g_{_B}g_{_{YB}}\frac{x\times tb'}{T}}\\ [6pt]
{-\frac{1}{4}g^2\frac{ x^2tb}{1+tb^2}}\hspace{-0.1cm}-\hspace{-0.1cm}n^2&{\frac{1}{4}\frac{g^2 x^2tb^2 }{1+tb^2}\hspace{-0.1cm}+\hspace{-0.1cm}\frac{n^2}{tb}}&
{\frac{1}{2}g_{_B}g_{_{YB}}\frac{x\times tb}{T}}&{\frac{1}{2}g_{_B}g_{_{YB}}\frac{x\times tb\times tb'}{T}}\\ [6pt]
{\frac{1}{2}g_{_B}g_{_{YB}}\frac{x}{T}}&{\frac{1}{2}g_{_B}g_{_{YB}}\frac{x\times tb}{T}}&{\frac{g_{_B}^2}{1+tb'^2}\hspace{-0.1cm}+\hspace{-0.1cm}tb'N^2}&
{-g_{_B}^2\frac{tb'}{1+tb'^2}\hspace{-0.1cm}-\hspace{-0.1cm}N^2}\\ [6pt]
{-\frac{1}{2}g_{_B}g_{_{YB}}\frac{x\times tb'}{T}}&{\frac{1}{2}g_{_B}g_{_{YB}}\frac{x\times tb\times tb'}{T}}&
{-g_{_B}^2\frac{tb'}{1+tb'^2}\hspace{-0.1cm}-\hspace{-0.1cm}N^2}&{g_{_B}^2\frac{tb'^2}{1+tb'^2}\hspace{-0.1cm}+\hspace{-0.1cm}\frac{N^2}{tb'}}
\end{array}\right).
\end{eqnarray}
Here, $g^2=g_{_1}^2+g_{_2}^2+g_{_{YB}}^2$, $T=\sqrt{1+tb^2}\sqrt{1+tb'^2}$,
$n^2=\frac{{\rm Re}B\mu}{u^2}$ and $N^2=\frac{{\rm Re}B_\eta}{u^2}$.

Then, we will consider the radiative corrections $\Delta \Pi$ from the one-loop zero temperature effective potential to the tree level Higgs mass squared matrix.
\begin{eqnarray}
m_h^2=M_h^2+\Delta\Pi,~~~\Delta\Pi=\left(\begin{array}{*{20}{c}}
\Delta\Pi_{11}&\Delta\Pi_{12}&\Delta\Pi_{13}&\Delta\Pi_{14}\\ \Delta\Pi_{21}&\Delta\Pi_{22}&\Delta\Pi_{23}&\Delta\Pi_{24}\\
\Delta\Pi_{31}&\Delta\Pi_{32}&\Delta\Pi_{33}&\Delta\Pi_{34}\\ \Delta\Pi_{41}&\Delta\Pi_{42}&\Delta\Pi_{43}&\Delta\Pi_{44}\end{array}\right),
\end{eqnarray}
with $\Delta\Pi_{ii}=[-\frac{1}{\phi_i}\frac{\partial \Delta V_1}{\partial \phi_i}+\frac{\partial^2 \Delta V_1}{\partial \phi_i^2}]_{\phi_i=v_i}$ and $\Delta\Pi_{ij}=[\frac{\partial^2 \Delta V_1}{\partial \phi_i\partial \phi_j}]_{\phi_i=v_i,\phi_j=v_j}$, $v_i,v_j\in(v_d,v_u,v_\eta,v_{\bar{\eta}}$) and $\phi_{i,j}$ are the CP-even Higgs components $\phi_d,\;\phi_u,\;\phi_{\eta},\;\phi_{\bar{\eta}}$.
$\Delta V_1$ represents the one-loop zero temperature effective potential whose full form has be discussed in the literature\cite{effpotential0}. In principle, the radiative correction is dominated by the contributions of top quark, bottom quark, stop quarks and sbottom quarks.
\begin{eqnarray}
&&\Delta V_1=-\frac{3}{({4\pi})^2}[\frac{m^4_{t}}{4}(\ln(\frac{m^2_{t}}{\lambda^2})-\frac{3}{2})
-\frac{1}{2}\frac{m^4_{\tilde{t}}}{4}(\ln(\frac{m^2_{\tilde{t}}}{\lambda^2})-\frac{3}{2})]
\nonumber\\&&\hspace{1.3cm}-\frac{3}{({4\pi})^2}[\frac{m^4_{b}}{4}(\ln(\frac{m^2_{b}}{\lambda^2})-\frac{3}{2})
-\frac{1}{2}\frac{m^4_{\tilde{b}}}{4}(\ln(\frac{m^2_{\tilde{b}}}{\lambda^2})-\frac{3}{2})].
\end{eqnarray}
Here, the masses of top quark and bottom quark are respectively $m_t=\frac{1}{\sqrt{2}}Y_t\phi_u$ and $m_b=\frac{1}{\sqrt{2}}Y_b\phi_d$. Additionally, the mass squared matrix of stop quark and sbottom quark will be deduced respectively in the basis $(\tilde U_L, \tilde U_R)$ and $(\tilde D_L, \tilde D_R)$.
\begin{eqnarray}
m_{\tilde t}^2=
\left(\begin{array}{cc}m^2_{\tilde{t}_L}&m^2_{\tilde{t}_{LR}}\\
m^2_{\tilde{t}_{RL}}&m^2_{\tilde{t}_R}\end{array}\right),~~~~~m_{\tilde b}^2=
\left(\begin{array}{cc}m^2_{\tilde{b}_L}&m^2_{\tilde{b}_{LR}}\\
m^2_{\tilde{b}_{RL}}&m^2_{\tilde{b}_R}\end{array}\right),
\end{eqnarray}
where,
\begin{eqnarray}
&&m^2_{\tilde{t}_L}= +\frac{1}{24} \Big[3 g_{2}^{2} \Big(- \phi_{u}^{2}  + \phi_{d}^{2}\Big)+\Big(g_{1}^{2} + g_{Y B}^{2}\Big)\Big(- \phi_{d}^{2}  + \phi_{u}^{2}\Big)-2 g_{B}^{2}\Big(- \phi_{\bar{\eta}}^{2}  + \phi_{\eta}^{2}\Big)\nonumber\\
&&\hspace{1.3cm}+g_{Y B} g_{B} \Big(2 \phi_{\bar{\eta}}^{2}  -2 \phi_{\eta}^{2}  - \phi_{d}^{2}  + \phi_{u}^{2}\Big)\Big] +\frac{1}{2} \Big(2 m_{\tilde{Q}_3}^2  + \phi_{u}^{2} {Y_{t}^{\dagger}  Y_t} \Big),\nonumber\\
&&m^2_{\tilde{t}_R}= +\frac{1}{24} \Big[2 g_{B}^{2} \Big(- \phi_{\bar{\eta}}^{2}  + \phi_{\eta}^{2}\Big) + 4 \Big(g_{1}^{2} + g_{Y B}^{2}\Big)\Big(- \phi_{u}^{2}  + \phi_{d}^{2}\Big) \nonumber\\
&&\hspace{1.3cm}+  g_{Y B} g_{B}\Big(-8 \phi_{\bar{\eta}}^{2}  + 8 \phi_{\eta}^{2}  - \phi_{u}^{2}  + \phi_{d}^{2}\Big)\Big]+\frac{1}{2} \Big(2 m_{\tilde{U}_3}^2  + \phi_{u}^{2} {Y_t  Y_{t}^{\dagger}} \Big),\nonumber\\
&&m^2_{\tilde{t}_{RL}}=(m^2_{\tilde{t}_{LR}})^{\dagger}=\frac{1}{\sqrt2}(\phi_u T_{t}-\phi_d\mu^* Y_{t)}.
\end{eqnarray}
Similarly, the elements in the sbottom mass squared matrix are
\begin{eqnarray}
&&m^2_{\tilde{b}_L}= +\frac{1}{24}\Big[-2 g_{B}^{2} \Big(- \phi_{\bar{\eta}}^{2}  + \phi_{\eta}^{2}\Big) + \Big(3 g_{2}^{2}  + g_{1}^{2} + g_{Y B}^{2}\Big)\Big(- \phi_{d}^{2}  + \phi_{u}^{2}\Big) \nonumber \\
&&\hspace{1.3cm}+  g_{Y B} g_{B} \Big(2 \phi_{\bar{\eta}}^{2}  -2 \phi_{\eta}^{2}  - \phi_{d}^{2}  + \phi_{u}^{2}\Big)\Big]+\frac{1}{2} \Big(2 m_{\tilde{Q}_3}^2  + \phi_{d}^{2}{Y_{b}^{\dagger}Y_b}\Big),\nonumber\\
&&m^2_{\tilde{b}_R}= +\frac{1}{24} \Big[2 \Big(\Big(g_{1}^{2} + g_{Y B}^{2}\Big)\Big(- \phi_{d}^{2}  + \phi_{u}^{2}\Big) + g_{B}^{2} \Big(- \phi_{\bar{\eta}}^{2}  + \phi_{\eta}^{2}\Big)\Big) \nonumber \\
&&\hspace{1.3cm}+   g_{Y B} g_{B} \Big(4 \phi_{\bar{\eta}}^{2}  -4 \phi_{\eta}^{2}  - \phi_{u}^{2}  + \phi_{d}^{2}\Big)\Big]+\frac{1}{2} \Big(2 m_{\tilde{D}_3}^2  + \phi_{d}^{2} {Y_b  Y_{b}^{\dagger}} \Big),\nonumber \\
&&m^2_{\tilde{b}_{RL}}=(m^2_{\tilde{b}_{LR}})^{\dagger}=\frac{1}{\sqrt{2}}\Big(\phi_d T_b  - \phi_u Y_b \mu^* \Big).
\end{eqnarray}
Then, the mass eigenvalues of stop and sbottom quarks will be given by
\begin{eqnarray}
&&m^2_{\tilde{t}_{1,2}}=\frac{1}{2}(m^2_{\tilde{t}_{L}}+ m^2_{\tilde{t}_{R}})\pm
\frac{1}{2}[(m^2_{\tilde{t}_{L}}-m^2_{\tilde{t}_{R}})^2+4m^2_{\tilde{t}_{LR}}m^2_{\tilde{t}_{RL}}]^\frac{1}{2},\nonumber\\
&&m^2_{\tilde{b}_{1,2}}=\frac{1}{2}(m^2_{\tilde{b}_{L}}+ m^2_{\tilde{b}_{R}})\pm
\frac{1}{2}[(m^2_{\tilde{b}_{L}}-m^2_{\tilde{b}_{R}})^2+4m^2_{\tilde{b}_{LR}}m^2_{\tilde{b}_{RL}}]^\frac{1}{2}.
\end{eqnarray}

Therefore, the radiative corrections $\Delta \Pi$ from the one-loop zero temperature effective potential will be deduced as:
\begin{eqnarray}
&&\Delta \Pi_{ii}=\frac{3}{(4\pi)^2}\Big[\sum_{k=t,b}\Big(\frac{2}{\phi_i}f(m^2_k)\frac{\partial m^2_k}{\partial \phi_i}-2f(m^2_k)\frac{\partial^2 m^2_k}{\partial \phi^2_i}-\frac{1}{2}\ln(\frac{m^2_k}{\Lambda^2})\frac{\partial m^2_k}{\partial \phi_i}\frac{\partial m^2_k}{\partial \phi_i}\Big)\nonumber \\
&&\hspace{1.5cm}+\sum_{k=\tilde{t}_1,\tilde{t}_2,\tilde{b}_1,\tilde{b}_2}\Big(-\frac{1}{\phi_i}f(m^2_k)\frac{\partial m^2_k}{\partial \phi_i}+f(m^2_{k})\frac{\partial^2 m^2_{k}}{\partial \phi_i^2}+\frac{1}{4}\ln(\frac{m^2_{k}}{\Lambda^2})\frac{\partial m^2_{k}}{\partial \phi_i}\frac{\partial m^2_{k}}{\partial \phi_i}\Big)\Big],\nonumber \\
&&\Delta \Pi_{ij}=\frac{3}{(4\pi)^2}\sum_{k=\tilde{t}_1,\tilde{t}_2,\tilde{b}_1,\tilde{b}_2}\Big(f(m^2_{k})\frac{\partial^2 m^2_{k}}{\partial \phi_i\partial \phi_j}+\frac{1}{4}\ln(\frac{m^2_{k}}{\Lambda^2})\frac{\partial m^2_{k}}{\partial \phi_i}\frac{\partial m^2_{k}}{\partial \phi_j}\Big).
\end{eqnarray}
Here, $f(m_k^2)=\frac{1}{4}m_k^2[\ln(\frac{m_k^2}{\Lambda^2})-1]$, $\Lambda$ is the new physics scale and we take $\Lambda=1 ~\rm TeV$ in the following numerical discussion. $m_k$ represent the corresponding particle masses. The square matrix $m_h^2$ will be diagonalized to the mass eigenstate by the unitary matrix $Z_{h_i}$.
\subsection{Numerical discussion of the Higgs mass in the B-LSSM}
The mass of physical Higgs boson reads $m_{h^0}=125.1\pm0.14~\rm GeV$ by the latest LHC experiments\cite{PDG}. The updated experimental data on the mass of $Z'$ boson indicates $M'_Z > 4.05~ {\rm TeV}$ with $95\%$ confidence level(CL). Refs.\cite{Zpupper1,Zpupper2} give us an upper bound on the ratio between the mass of $Z'$ boson and its gauge coupling at $99\%$ CL as $\frac{M'_Z}{g_B}\geq 6~{\rm TeV}$. We choose $M'_Z = 4.2~ {\rm TeV}$ in our numerical calculation, so the value of parameter $g_B$ is restricted in the region of $0 < g_B \leq0.7$. The Yukawa coupling $Y_b$, determined by the parameter $tb$, is defined as $Y_b=\sqrt{2(tb^2+1)}m_b/v$. In general, the value of $Y_b$ is smaller than 1, $m_b\simeq4.18~\rm GeV$ and $v\simeq246~\rm GeV$, so the parameter $tb$ should be approximatively smaller than 40. Besides, the large $tb$ has been excluded by the $\bar{B}\rightarrow X_s\gamma$ experiment. The coupling parameter $g_{YB}$ will be taken around $-0.45\sim-0.05$, and the reason why constant $g_{YB}$ is negative has been discussed specifically in Ref.\cite{gYB}. In addition, LHC searches constrain $tb'<1.5$\cite{tbB}.
\begin{figure}[t]
\centering
\includegraphics[width=8cm]{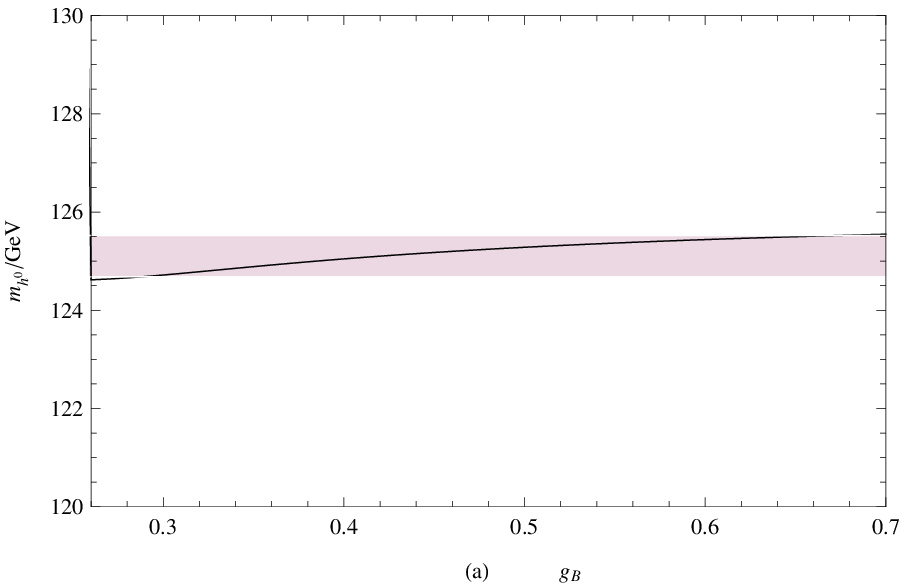}
\includegraphics[width=8cm]{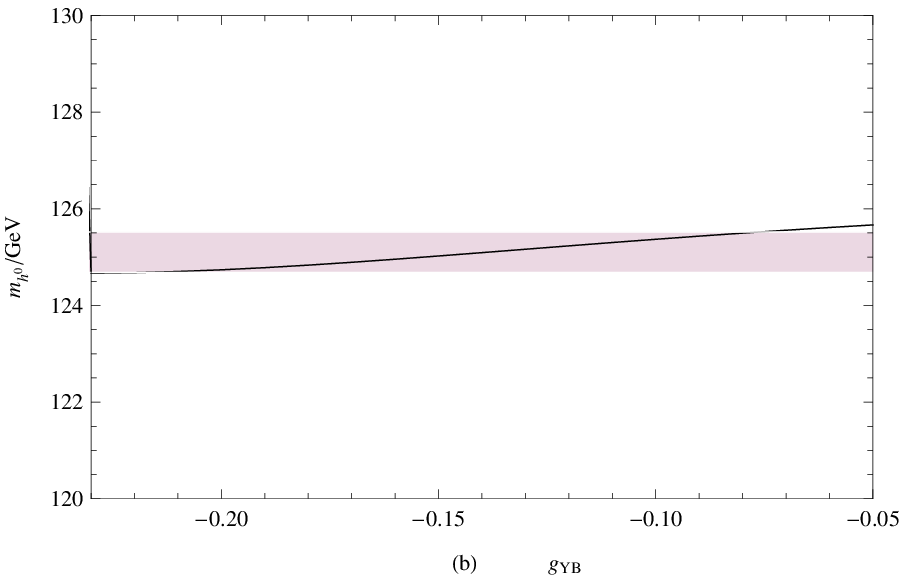}\\
\includegraphics[width=8cm]{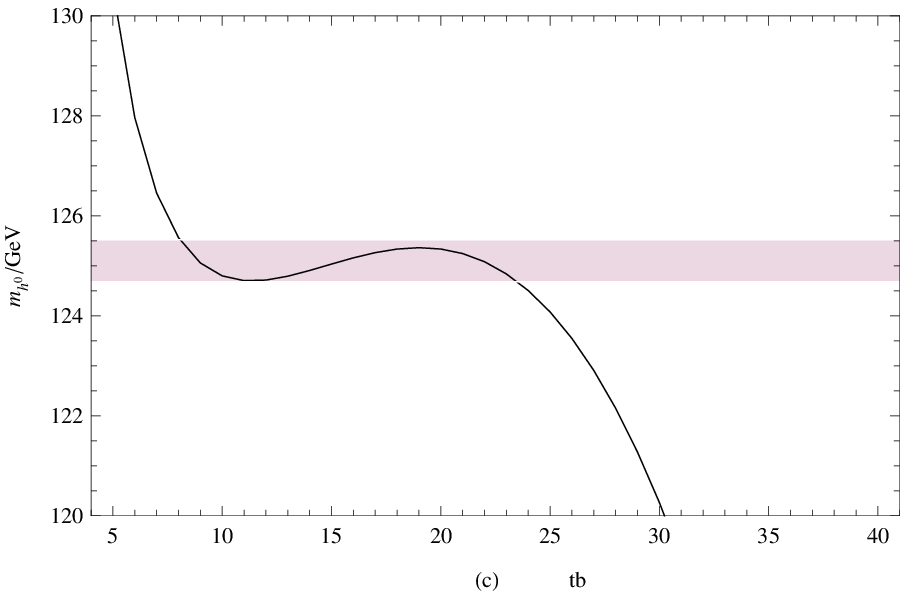}
\includegraphics[width=8cm]{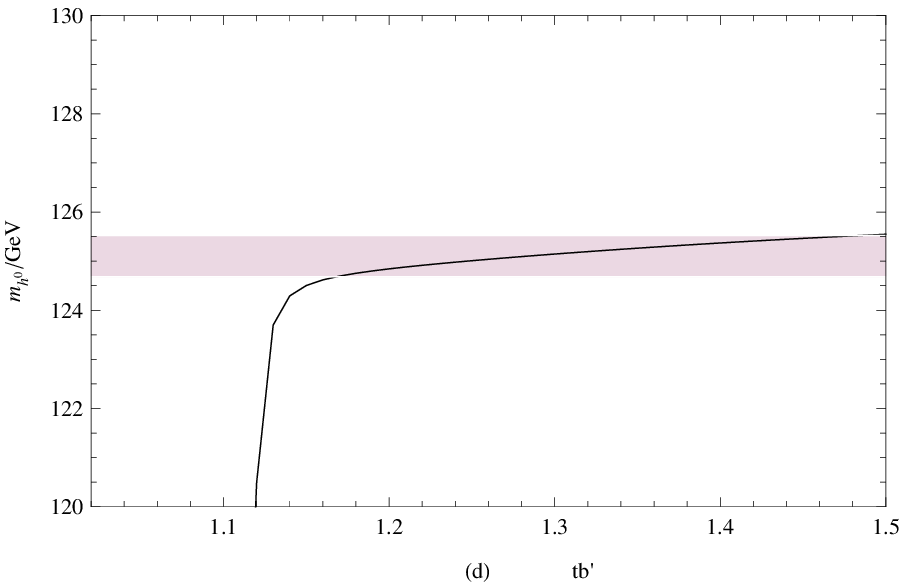}
\caption{The mass of physical Higgs boson changes with the parameters $g_B$, $g_{YB}$, $tb$ and $tb'$ respectively. The pink area denotes the experimental $3\sigma$ interval. }
\label{h0}
\end{figure}

First, we take $B_{\mu}=1.1~\rm TeV^2,~B_{\eta}=0.8~\rm TeV^2,~\mu=0.8~\rm TeV,~m_{\tilde{Q}_3}=m_{\tilde{U}_3}=3.6~\rm TeV,~m_{\tilde{D}_3}=2.5~\rm TeV,~T_t=0.6~\rm TeV,~T_b=1.7~\rm TeV$. The physical Higgs mass changing with parameter $g_B$ will be shown in FIG.\ref{h0}(a) with $tb=22,~tb'=1.4,~g_{YB}=-0.1$; The physical Higgs mass versus parameter $g_{YB}$ will be shown in FIG.\ref{h0}(b) with $tb=22,~tb'=1.4,~g_{B}=0.6$; The physical Higgs mass versus parameter $tb'$ will be plotted in FIG.\ref{h0}(d) with $tb=22,~g_{B}=0.6,~g_{YB}=-0.1$. When $tb'=1.46,~\mu=1~\rm TeV,~m_{\tilde{U}_3}=3.4~\rm TeV,~m_{\tilde{D}_3}=2.5~\rm TeV,$ and $T_t=0.45~\rm TeV$, the physical Higgs mass affected by parameter $tb$ will be searched in FIG.\ref{h0}(c). As shown in picture, the physical Higgs mass slightly increases with the enlarging $g_B$, which will be limited in the region $0.3\sim0.7$ to obtain suitable Higgs mass. When the physical Higgs mass satisfies experimental $3\sigma$ interval, parameters $g_{YB},~tb$ and $tb'$ will be constrained within the small regions: $-0.23\leq g_{YB}\leq-0.07$, $7\leq tb\leq25$ and $1.15\leq tb'<1.5$. So $g_{YB},~tb$ and $tb'$ are all sensitive parameters. Besides, the physical Higgs mass is mainly determined by the $\phi_u$ component, which can be up to $99.8\%$ contributions.
\begin{figure}[t]
\centering
\includegraphics[width=8cm]{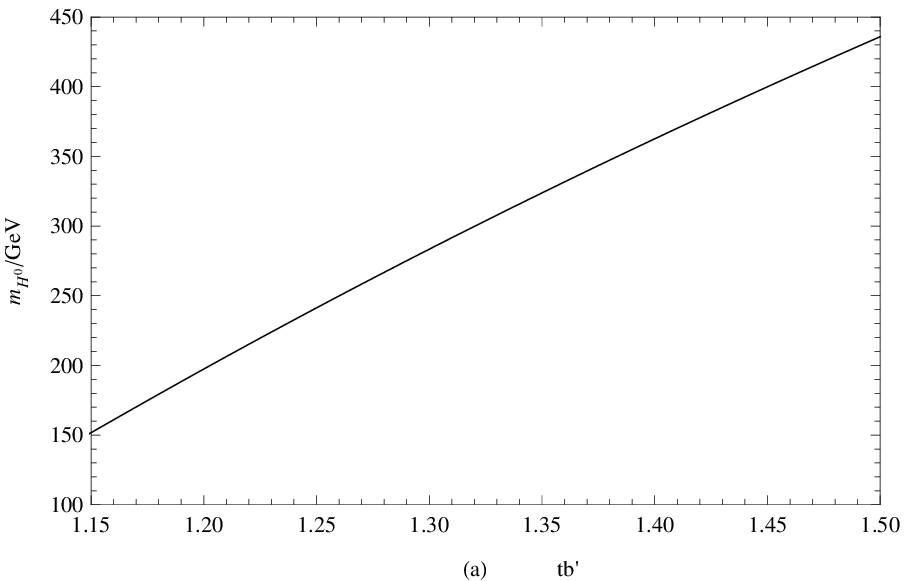}
\includegraphics[width=8cm]{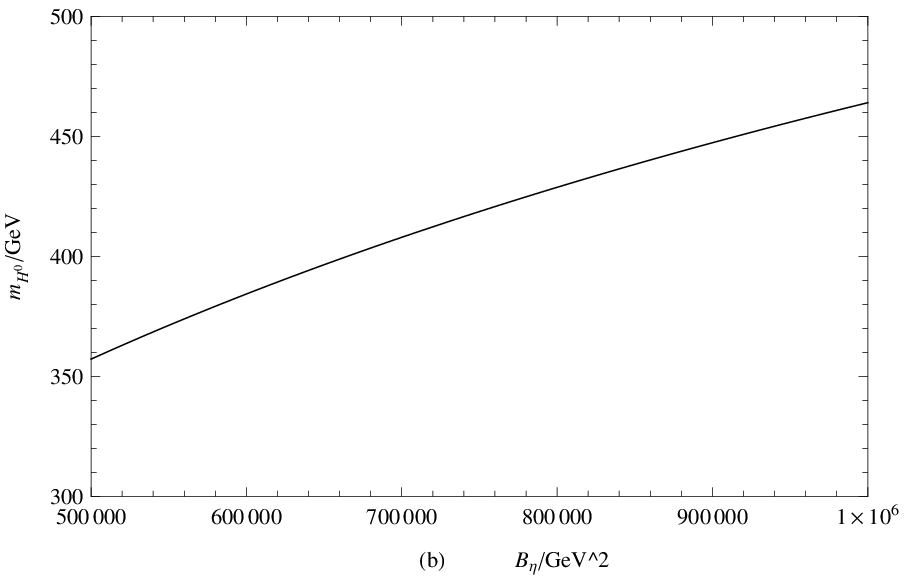}
\caption{The mass of second-light Higgs boson changes with the parameters $B_{\eta}$ and $tb'$ respectively. }
\label{H0}
\end{figure}

Then, The second-light Higgs mass $m_{H^0}$ will be discussed in FIG.\ref{H0} with parameters $tb'$ and $B_{\eta}$. As $tb=22$, the $m_{H^0}$ versus $tb'$ will be described in FIG.\ref{H0}(a). We can find that the second-light Higgs mass increases quickly with the increasing $tb'$. So parameter $tb'$ affects the second-light Higgs mass smartly. In FIG.\ref{H0}(b), the $m_{H^0}$ changing with $B_{\eta}$ will be studied with $tb=23$ and $tb'=1.49$. The larger $B_{\eta}$, the larger $m_{H^0}$ we can obtain. However, the influence from parameter $B_{\eta}$ on the second-light Higgs mass is gentle.
\section{finite temperature effective potential in the B-LSSM}
The value of $u$ in the $U(1)_{B-L}$ Higgs singlets is around $10~\rm{TeV}$, which is much larger than $v\simeq246~\rm{GeV}$ of Higgs doublets. When the temperature of the universe is higher than the scale of the new physics, both $U(1)_{B-L}$ Higgs singlets and Higgs doublets are supposed to be trapped at the origins of their potential. As the temperature drops down to the scale of the new physics, the first-order phase transition with the two Higgs singlets may occur because the scale of Higgs potential becomes the new physics one. So we ignore two Higgs doublets part and their interaction terms in the following theoretical analyses and numerical discussions. Then, we consider the tree-level scalar potential including the two singlet superfields $\hat{\eta}$ and $\hat{\bar{\eta}}$, which will be written as
\begin{eqnarray}
&&\hspace{-0.6cm}V_0=\frac{1}{2}\big[g_B(\frac{1}{6}Q^{I*}Q^I-\frac{1}{6}D^{I*}D^I-\frac{1}{6}U^{I*}U^I
-\frac{1}{2}L^{I*}L^I+\frac{1}{2}E^{I*}E^I+\frac{1}{2}\nu^{I*}\nu^I-\eta^\dag\eta+\bar{\eta}^\dag\bar{\eta})\nonumber\\
&&\hspace{-0.1cm}+\frac{1}{2}g_{YB}(\frac{1}{3}Q^{I*}Q^I+\frac{2}{3}D^{I*}D^I-\frac{4}{3}U^{I*}U^I
-L^{I*}L^I+2E^{I*}E^I)\big]^2+|\mu_{\eta}|^2\eta^\dag\eta+|\mu_{\eta}|^2\bar{\eta}^\dag\bar{\eta}\nonumber\\
&&\hspace{-0.1cm}-(\mu_{\eta}Y_x\bar{\eta}^\dag\nu^{I}\nu^I+{\rm H.c.})+Y_x^2|\nu^{I}\nu^I|^2+4Y_x^2|\eta\nu^I|^2
+m_{\eta}^2\eta^\dag\eta+m_{\bar{\eta}}^2\bar{\eta}^\dag\bar{\eta}-(B_{\eta}\eta\bar{\eta}+{\rm H.c.}).
\end{eqnarray}
Here, the contribution of $-B_{\eta}\eta\bar{\eta}$ is deduced from the softbreaking terms, and parameter $B_{\eta}$ possesses mass square dimension..

\subsection{Finite temperature effective potential}
The GW spectrum is generated by the first-order PT, which is determined by the finite temperature effective potential. In the B-LSSM, the finite temperature effective potential for both zero and finite temperatures are essential for realizing the first-order PT. We are actually interested in the ${\rm Re}\eta\equiv\frac{1}{\sqrt{2}}\phi_{\eta}$ and ${\rm Re}\bar{\eta}\equiv\frac{1}{\sqrt{2}}\phi_{\bar{\eta}}$. Therefore, the $(\phi_{\eta},\phi_{\bar{\eta}})$-dependent finite temperature effective potential can be expressed as
\begin{eqnarray}
V_{eff}(\phi_{\eta},\phi_{\bar{\eta}},T)=V_0(\phi_{\eta},\phi_{\bar{\eta}})+\Delta V_1(\phi_{\eta},\phi_{\bar{\eta}},0)+\Delta V_1(\phi_{\eta},\phi_{\bar{\eta}},T)+\Delta V_{daisy}(\phi_{\eta},\phi_{\bar{\eta}},T),
\end{eqnarray}
where $V_0(\phi_{\eta},\phi_{\bar{\eta}})$ and $\Delta V_1(\phi_{\eta},\phi_{\bar{\eta}},0)$ represent the tree-level and one-loop zero temperature effective potential respectively. $\Delta V_1(\phi_{\eta},\phi_{\bar{\eta}},T)$ are the one-loop finite temperature contributions, and $\Delta V_{daisy}(\phi_{\eta},\phi_{\bar{\eta}},T)$ are the daisy corrections. The concrete expressions will be concluded as\cite{effpotential1,effpotential2}
\begin{eqnarray}
&&V_0(\phi_{\eta},\phi_{\bar{\eta}})=\frac{1}{2}\Big[\frac{1}{4}g_B^2(\phi_{\eta}^*\phi_{\eta}-\phi_{\bar{\eta}}^*\phi_{\bar{\eta}}
)^2+(|\mu_{\eta}|^2+m_{\eta}^2)\phi_{\eta}^*\phi_{\eta}\nonumber\\
&&\hspace{1.9cm}+(|\mu_{\eta}|^2+m_{\bar{\eta}}^2)\phi_{\bar{\eta}}^*\phi_{\bar{\eta}}
+(-B_{\eta}\phi_{\eta}\phi_{\bar{\eta}}+{\rm H.c.})\Big],\nonumber\\
&&\Delta V_1(\phi_{\eta},\phi_{\bar{\eta}},0)=\sum_i(-1)^k\frac{n_i}{64\pi^2}m_i^4(\phi_{\eta},\phi_{\bar{\eta}})
\Big[\log\frac{m_i^2(\phi_{\eta},\phi_{\bar{\eta}})}{\Lambda^2}-c_i\Big],\nonumber\\
&&\Delta V_1(\phi_{\eta},\phi_{\bar{\eta}},T)=\frac{T^4}{2\pi^2}\sum_i(-1)^k n_i J_i\Big[\frac{m_i^2(\phi_{\eta},\phi_{\bar{\eta}})}{T^2}\Big],\nonumber\\
&&\Delta V_{daisy}(\phi_{\eta},\phi_{\bar{\eta}},T)=-\frac{T}{12\pi}\sum_{i=boson} n_i[{\cal M}_i^3(\phi_{\eta},\phi_{\bar{\eta}},T)-m_i^3(\phi_{\eta},\phi_{\bar{\eta}})].
\end{eqnarray}
$m_i^2$ are the $(\phi_{\eta},\phi_{\bar{\eta}})$-dependent particle masses square of fermions and bosons. $n_i$ are the degrees of freedom corresponding to the $(\phi_{\eta},\phi_{\bar{\eta}})$-dependent particles. $k=1(0)$ and $c_i=\frac{3}{2}(\frac{5}{6})$ correspond to fermions (bosons). The thermal function corresponding to boson and fermion particles can be written as $J_{boson}(x^2)=\int_0^{\infty}dy y^2\log(1-\exp[y^2+x^2])$ and $J_{fermion}(x^2)=\int_0^{\infty}dy y^2\log(1+\exp[y^2+x^2])$. The masses square ${\cal M}_i^2$ are obtained from the $m_i^2$ by adding the T-dependent self-energy corrections\cite{Tcorrection1,Tcorrection2,Tcorrection3,Tcorrection4}. We will discuss the $(\phi_{\eta},\phi_{\bar{\eta}})$-dependent particle masses square $m_i^2$ and ${\cal M}_i^2$ specifically as follows.

The $(\phi_{\eta},\phi_{\bar{\eta}})$-dependent masses square of the CP-even Higgs and CP-odd Higgs are shown as
\begin{eqnarray}
&&m_{\phi_{\eta}\phi_{\eta}}^2=g_B^2\phi_{\eta}^2+B_{\eta}\frac{\phi_{\bar{\eta}}}{\phi_{\eta}},
m_{\phi_{\bar{\eta}}\phi_{\bar{\eta}}}^2=g_B^2\phi_{\bar{\eta}}^2+B_{\eta}\frac{\phi_{\eta}}{\phi_{\bar{\eta}}},\nonumber\\
&&m_{\sigma_{\eta}\sigma_{\eta}}^2=B_{\eta}\frac{\phi_{\bar{\eta}}}{\phi_{\eta}},
m_{\sigma_{\bar{\eta}}\sigma_{\bar{\eta}}}^2 =B_{\eta}\frac{\phi_{\eta}}{\phi_{\bar{\eta}}}.
\end{eqnarray}
The $(\phi_{\eta},\phi_{\bar{\eta}})$-dependent masses square of the squarks and sleptons, CP-even and CP-odd sneutrinos can be written as
\begin{eqnarray}
&&m_{\tilde{Q}^I}^2 =-\frac{1}{12}(g_B^2+g_Bg_{YB})(\phi_{\eta}^2-\phi_{\bar{\eta}}^2),
m_{\tilde{U}^I}^2 =\frac{1}{12}(g_B^2+4g_Bg_{YB})(\phi_{\eta}^2-\phi_{\bar{\eta}}^2),\nonumber\\
&&m_{\tilde{D}^I}^2 =\frac{1}{12}(g_B^2-2g_Bg_{YB})(\phi_{\eta}^2-\phi_{\bar{\eta}}^2);
\end{eqnarray}
\begin{eqnarray}
&&m_{\tilde{L}^I}^2 =\frac{1}{4}(g_B^2+g_Bg_{YB})(\phi_{\eta}^2-\phi_{\bar{\eta}}^2),
m_{\tilde{R}^I}^2 =-\frac{1}{4}(g_B^2+2g_Bg_{YB})(\phi_{\eta}^2-\phi_{\bar{\eta}}^2),\nonumber\\
&&m_{\tilde{\nu}_i}^2 =m_{\tilde{\nu}_i^R}^2 =-\frac{1}{4}g_B^2(\phi_{\eta}^2-\phi_{\bar{\eta}}^2).
\end{eqnarray}

In the B-LSSM, there is a $(\phi_{\eta},\phi_{\bar{\eta}})$-dependent $U(1)_{B-L}$ gauge boson, whose mass square is given by
\begin{eqnarray}
m_{Z'}^2 =g_B^2(\phi_{\eta}^2+\phi_{\bar{\eta}}^2).
\end{eqnarray}

In the basis $(\tilde{B}',\tilde{\eta},\tilde{\bar{\eta}})$, we obtain the neutralino mass matrix, which is a Majorana fermionic component.
\begin{equation}
m_{\tilde{\chi}^0} = \left(
\begin{array}{ccc}
0&- g_{B} \phi_{\eta}  &g_{B} \phi_{\bar{\eta}} \\
- g_{B} \phi_{\eta}  &0 &0 \\
g_{B} \phi_{\bar{\eta}}  &0  &0\end{array}
\right).
\end{equation}
Diagonalizing the $m_{\tilde{\chi}^0}^{\dag}m_{\tilde{\chi}^0}$, we can obtain the corresponding mass eigenvalues, which are $g_B^2(\phi_{\eta}^2+\phi_{\bar{\eta}}^2)$, $g_B^2(\phi_{\eta}^2+\phi_{\bar{\eta}}^2)$ and 0.

Then, considering the $T^2$-proportional part of the one-loop corrections, we adopt the ${\cal M}_i^2$ for the Higgs, squark, slepton, sneutrino and $Z'$ boson. Here, the temperature corrections for the Higgs boson and the third squark consider the Yukawa couplings $Y_t$ and $Y_b$, which are both determined by the parameter $tb$. Besides, we only consider the temperature contributions to the longitudinal component of $Z'$ boson.
\begin{eqnarray}
&&{\cal M}_{\phi_{\eta}\phi_{\eta}}^2= m_{\phi_{\eta}\phi_{\eta}}^2+\frac{T^2}{2}g_B^2\;,\;{\cal M}_{\phi_{\bar{\eta}}\phi_{\bar{\eta}}}^2= m_{\phi_{\bar{\eta}}\phi_{\bar{\eta}}}^2+\frac{T^2}{2}g_B^2;\nonumber\\
&&{\cal M}_{\sigma_{\eta}\sigma_{\eta}}^2 = m_{\sigma_{\eta}\sigma_{\eta}}^2+\frac{T^2}{2}g_B^2\;,\,
{\cal M}_{\sigma_{\bar{\eta}}\sigma_{\bar{\eta}}}^2
=m_{\sigma_{\bar{\eta}}\sigma_{\bar{\eta}}}^2+\frac{T^2}{2}g_B^2;
\end{eqnarray}
\begin{eqnarray}
&&{\cal M}_{\tilde{Q}^{1,2}}^2 =m_{\tilde{Q}^{1,2}}^2+\frac{T^2}{8}(\frac{16}{3}g_s^2+3g_2^2+\frac{1}{3}g_1^2
+\frac{1}{36}(4g_B^2+4g_{YB}^2+7g_Bg_{YB})),\nonumber\\
&&{\cal M}_{\tilde{Q}^3}^2 =m_{\tilde{Q}^3}^2+\frac{T^2}{8}(\frac{16}{3}g_s^2+3g_2^2+\frac{1}{3}g_1^2+2(Y_t^2+Y_b^2)
+\frac{1}{36}(4g_B^2+4g_{YB}^2+7g_Bg_{YB})),\nonumber\\
&&{\cal M}_{\tilde{U}^{1,2}}^2 =m_{\tilde{U}^{1,2}}^2+\frac{T^2}{8}(\frac{16}{3}g_s^2+\frac{16}{9}g_1^2
+\frac{1}{9}(g_B^2+16g_{YB}^2+7g_Bg_{YB})),\nonumber\\
&&{\cal M}_{\tilde{U}^3}^2 =m_{\tilde{U}^3}^2 +\frac{T^2}{8}(\frac{16}{3}g_s^2+\frac{16}{9}g_1^2+4Y_t^2
+\frac{1}{9}(g_B^2+16g_{YB}^2+7g_Bg_{YB})),\nonumber\\
&&{\cal M}_{\tilde{D}^{1,2}}^2 =m_{\tilde{D}^{1,2}}^2 +\frac{T^2}{8}(\frac{16}{3}g_s^2+\frac{1}{3}g_1^2
+\frac{1}{9}(g_B^2+4g_{YB}^2+\frac{5}{2}g_Bg_{YB})),\nonumber\\
&&{\cal M}_{\tilde{D}^3}^2 =m_{\tilde{D}^3}^2 =+\frac{T^2}{8}(\frac{16}{3}g_s^2+\frac{1}{3}g_1^2+4Y_b^2
+\frac{1}{9}(g_B^2+4g_{YB}^2+\frac{5}{2}g_Bg_{YB}));
\end{eqnarray}
\begin{eqnarray}
&&{\cal M}_{\tilde{L}^I}^2 =m_{\tilde{L}^I}^2+\frac{T^2}{8}(3g_2^2+g_1^2
+\frac{1}{4}(4g_B^2+4g_{YB}^2+7g_Bg_{YB})),\nonumber\\
&&{\cal M}_{\tilde{R}^I}^2 =m_{\tilde{R}^I}^2+\frac{T^2}{8}(4g_1^2
+\frac{1}{4}(4g_B^2+7g_{YB}^2+14g_Bg_{YB})),\nonumber\\
&&{\cal M}_{\tilde{\nu}_i}^2 ={\cal M}_{\tilde{\nu}_i^R}^2 =m_{\tilde{\nu}_i}^2 +\frac{T^2}{8}g_B^2;
\end{eqnarray}
\begin{eqnarray}
{\cal M}_{Z'}^2 =m_{Z'}^2+\frac{T^2}{2}(2g_B^2+g_{YB}^2).
\end{eqnarray}
\section{GW generation by cosmological first-order phase transition}
The breaking of B-L symmetry not only induces first-order PT, but also induces second-order PT. The first-order PT is from a false vacuum state(electroweak symmetry phase) to a true one(electroweak symmetry broken phase), while the second-order PT is the process from an unstable state to a vacuum state.

The baryon asymmetry of the universe can be generated via the electroweak baryogenesis mechanism, which requires a strong first-order electroweak PT to provide a system deviation from thermal equilibrium. First-order PT will generate bubbles, inside which are electroweak symmetry broken phases, their externals are the electroweak symmetry phases. The universe begins with the electroweak symmetry phase, which is a metastable one. With the expansion of the bubbles, the collision and then fusion, the universe will reside at the electroweak symmetry broken phase. The GW will be generated along with the collision between bubbles, the plasma movement near the bubble wall disturbed by the expanding bubbles, and the magnetofluid turbulence caused by the plasma. The relative height of potential between false vacuum state and true one will determine the strength of GW. The GW can be detectable with the large enough relative height of potential, and the corresponding PT is called as the strong first-order PT.

The second-order PT can also generate bubbles initially. However, the background is unstable and will quickly roll to the vacuum state, so the bubbles and background will be exactly the same soon. The second-order PT will be more synchronous than the first-order one. Therefore, there is almost no bubble collision, plasma turbulence even the GW in the second-order PT. Therefore, we pay attention to the strong first-order PT to study the GW signals within the B-LSSM.
\subsection{Scalar potential parameters related to the GW spectrum}
In this section, we briefly discuss the properties of the GWs, which critically depend on two quantities: the ratio of the latent heat energy $\rho_{vac}$ to the radiation energy density $\rho_{rad}$ at the nucleation temperature $T_n$, which is defined as $\alpha$; $\beta$ is the speed of the PT at the nucleation temperature $T_n$. The transition from the symmetric phase to the symmetric broken phase takes place via thermal tunneling at the finite temperature. First-order PT pushes the bubbles of the symmetric broken phase to be nucleated, then they expand and eventually fill up the entire universe. The bubble nucleation rate per unit volume at the finite temperature is given by
\begin{eqnarray}
\Gamma\sim A(T)e^{-S}=A(T)e^{-S_E/T}.
\end{eqnarray}
$A(T)$ is a factor that is roughly proportional to $T^4$. $S$ represents the action in the four-dimensional Minkowski
space, while $S_E$ is the three-dimensional Euclidean action\cite{action1,action2}. Parameter $\beta$ can be defined as:
\begin{eqnarray}
\beta\equiv-\frac{dS}{dt}\Big|_{t=t_n}\simeq H_nT\frac{dS}{dT}\Big|_{T=T_n}=H_nT\frac{d(S_E/T)}{dT}\Big|_{T=T_n},
\end{eqnarray}
where $H_n$ denotes the Hubble rate at the nucleation temperature $T_n$. The key parameter that controls the GW signals is $\beta/{H_n}$. And the smaller $\beta/{H_n}$ will lead to the stronger PT and consequently the more sensitive GW signals.

Then, we will consider parameter $\alpha$, which denotes the strength of PT:
\begin{eqnarray}
\alpha\equiv\frac{\rho_{vac}}{\rho_{rad}},
\end{eqnarray}
where $\rho_{vac}$ represents the latent energy density released in the PT, while $\rho_{rad}$ denotes the radiation energy density.
\begin{eqnarray}
&&\rho_{vac}=\Big[\Delta V_{eff}(T)-T\frac{\partial\Delta V_{eff}(T)}{\partial T}\Big]\Big|_{T=T_n},\Delta V_{eff}(T)=V_{eff}(\phi_{sym},T)-V_{eff}(\phi_{bro},T),\nonumber\\
&&\rho_{rad}=\frac{\pi^2g_*}{30}T^4.
\end{eqnarray}
Here, $\phi_{sym}(\phi_{bro})$ releases the coordinate of the field at the symmetry(symmetric broken) phase. $g_*$ corresponds to the relativistic degree of freedom in the thermal plasma at $T_n$. We prefer to a larger value of $\alpha$, which will produce a much stronger PT, even a stronger GW spectrum.

\begin{figure}[t]
\centering
\includegraphics[width=5.3cm]{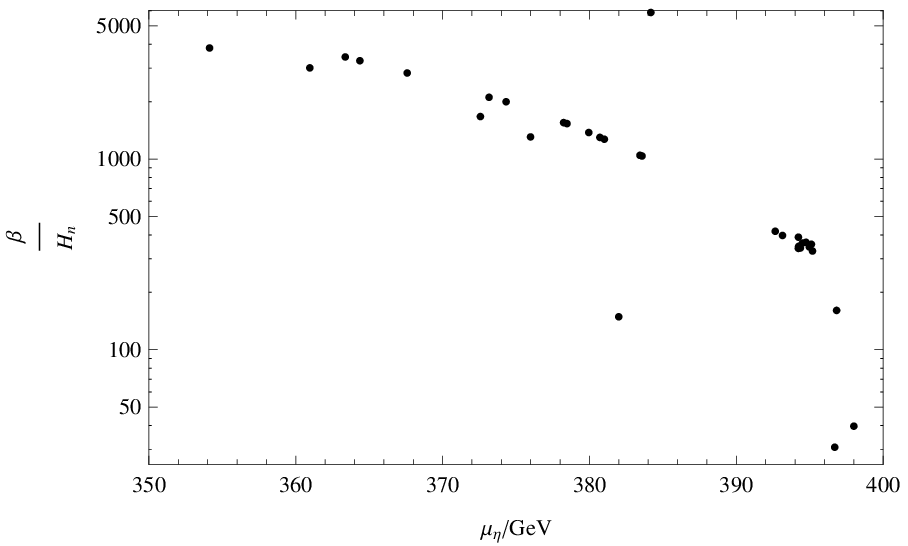}
\includegraphics[width=5.45cm]{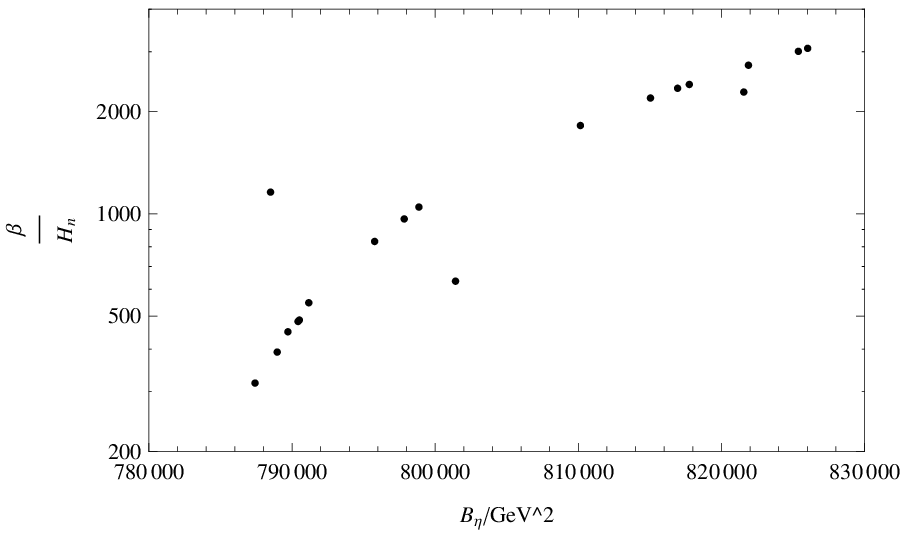}
\includegraphics[width=5.3cm]{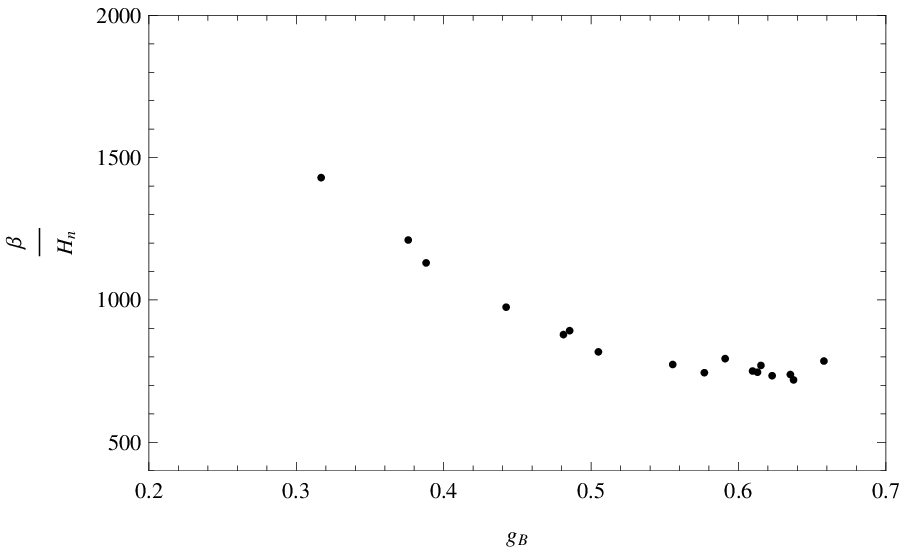}
\caption{The values of $\beta/{H_n}$ change with the parameters $\mu_{\eta}$, $B_{\eta}$ and $g_B$ respectively. }
\label{betaHn}
\end{figure}

\begin{figure}[t]
\centering
\includegraphics[width=8cm]{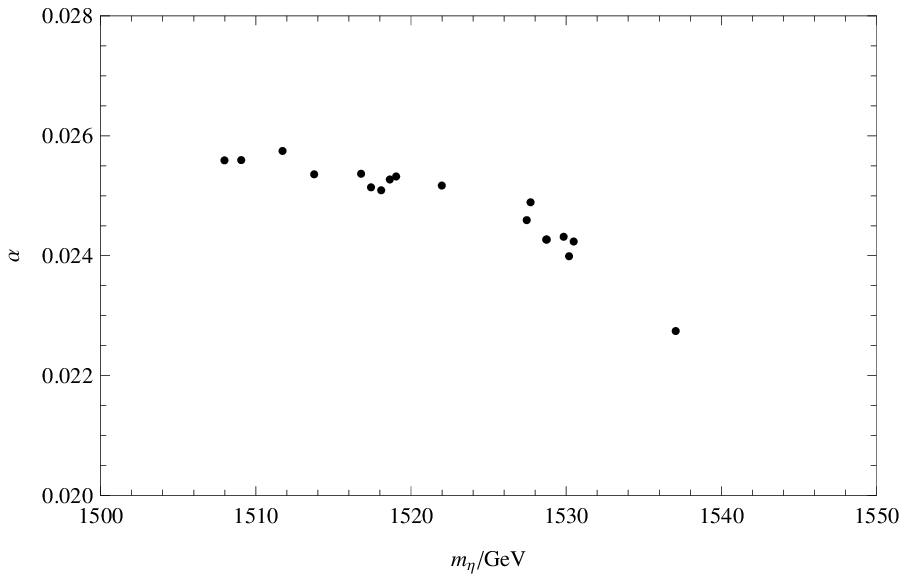}
\includegraphics[width=8cm]{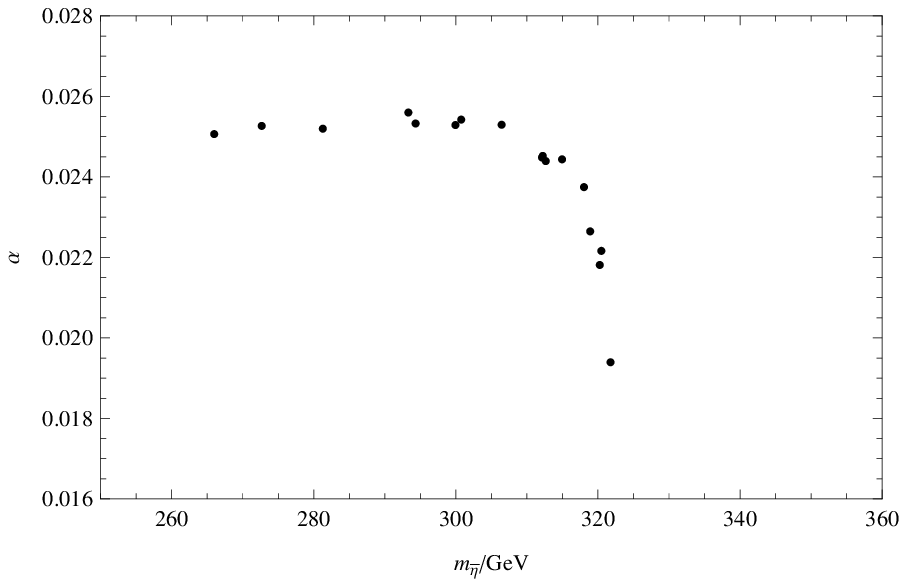}
\caption{The strengths of PT $\alpha$ change with the parameters $m_{\eta}$ and $m_{\bar{\eta}}$ respectively. }
\label{alpha}
\end{figure}

In FIG.\ref{betaHn}, we will discuss the $\beta/{H_n}$ changing with parameters $\mu_{\eta}$, $B_{\eta}$ and $g_B$ respectively. The values of $\beta/{H_n}$ decrease with the enlarging $\mu_{\eta}$ and $g_B$ respectively, while increase with the enlarging $B_{\eta}$.
The strength of PT $\alpha$ versus the parameters $m_{\eta}$ and $m_{\bar{\eta}}$ will be researched in FIG.\ref{alpha} respectively. With the enlarging $m_{\eta}$, the values of $\alpha$ decrease slowly. Besides, the values of $\alpha$ are around 0.025 as $m_{\bar{\eta}}$ in the region $260\sim310~\rm{GeV}$, and the value of $\alpha$ decreases with the gradually growing $m_{\bar{\eta}}$. We prefer to the smaller $\beta/{H_n}$ and larger $\alpha$ to obtain the more suitable GW signals. So, we will take $\mu_{\eta}=380~\rm{GeV}$, $B_{\eta}=8\times10^5~\rm{GeV^2}$, $g_B=0.6$, $m_{\eta}=1500~\rm{GeV}$ and $m_{\bar{\eta}}=300~\rm{GeV}$ in the following numerical discussion.
\subsection{GW spectrum}
In this section, we will discuss the GW spectrums specifically. There are three sources that generate the GW spectrums from the first-order PT: (1) The initial collision of scalar field and relevant shocks in the plasma. The technique of 'envelope approximation' has been widely used to model GW power spectrum from bubble collisions, which can be denoted by $h^2\Omega_{coll}$. (2) After the bubbles merging, the fluid kinetic energy waves in the plasma go on propagating outward into the broken phase. These waves spread at the sound speed in the plasma without the tractive force of scalar field bubble wall. Sound waves $h^2\Omega_{sw}$ are produced after the bubble collisions but before expansion. (3) The magnetohydrodynamic turbulent $h^2\Omega_{trub}$ of the surrounding plasma will be formed after the bubble collisions. These three contributions linearly combine, then we can obtain the corresponding GW spectrums, which will be expressed as
\begin{eqnarray}
h^2\Omega_{GW}\approx h^2\Omega_{coll}+h^2\Omega_{sw}+h^2\Omega_{trub}.
\label{GW}
\end{eqnarray}
\subsubsection{Bubble collisions}
Using the technique of 'envelope approximation', the contribution of the GWs generated from the bubble collisions is\cite{GWsource4,GWcollision}
\begin{eqnarray}
h^2\Omega_{coll}(f)= 1.65\times10^{-5}\kappa^2\Delta\Big(\frac{\beta}{H_n}\Big)^{-2}
\Big(\frac{\alpha}{1+\alpha}\Big)^2\Big(\frac{g_*}{100}\Big)^{-1/3}\frac{(a + b)(f/f_{coll})^a}{b + a(f/f_{coll})^{(a + b)}},
\label{GWcollf}
\end{eqnarray}
where $a=2.8$ and $b=1.0$. $\kappa$ denotes the efficiency factor of the latent heat deposited into a thin shell with $A=0.715$. The concrete expression of $\kappa$ and $\Delta$ will be discussed as follows\cite{GWsource4}
\begin{eqnarray}
\kappa = \frac{1}{1 + A\alpha}\Big(A\alpha + \frac{4}{27}\sqrt{\frac{3\alpha}{2}}\Big)\;,\;\;\;\Delta=\frac{0.11v_b^3}{0.42 + v_b^2},
\end{eqnarray}
here, $v_b=0.6$ characterizes the bubble wall velocity. The peak frequency $f_{coll}$ will be determined by the characteristic time-scale of the PT. From simulations, the peak frequency at $T_n$ is approximately given by $\frac{f_n}{\beta}=\frac{0.62}{1.8 - 0.1v_b + v_b^2}$, which is then red-shifted to yield the peak frequency today: $f_{coll}=1.67\times10^{-5}\Big(\frac{f_n}{\beta}\Big)\Big(\frac{\beta}{H_n}\Big)
\Big(\frac{T_n}{100\;{\rm GeV}}\Big)\Big(\frac{g_*}{100}\Big)^{1/6}\;{\rm Hz}$.
\subsubsection{Sound waves}
As a more significant and long-lasting source of GW, sound wave is produced by expanding sound shells in the fluid kinetic energy after the bubble collisions. The concrete sound wave contribution to the GWs is given by\cite{GWSW1,GWSW2}
\begin{eqnarray}
&&h^2\Omega_{sw}(f)= h^2\Omega_{sw}(f_{sw})\Big(\frac{f}{f_{sw}}\Big)^3\Big(\frac{7}{4+ 3(f/f_{sw})^2}\Big)^{7/2},
\label{GWswf}
\end{eqnarray}
where the concrete peak amplitude $h^2\Omega_{sw}(f_{sw})$ and the corresponding peak frequency$f_{sw}$ from sound waves can be shown as
\begin{eqnarray}
&&h^2\Omega_{sw}(f_{sw})=2.65\times10^{-6}\kappa_v^2v_b\Big(\frac{\beta}{H_n}\Big)^{-1}
\Big(\frac{\alpha}{1+\alpha}\Big)^2\Big(\frac{g_*}{100}\Big)^{-1/3},\nonumber\\
&&f_{sw}=1.9\times10^{-5}\frac{1}{v_b}\Big(\frac{\beta}{H_n}\Big)
\Big(\frac{T_n}{100\;{\rm GeV}}\Big)\Big(\frac{g_*}{100}\Big)^{1/6}\;{\rm Hz}.
\label{GWswfsw}
\end{eqnarray}
We take $\kappa_v\simeq\frac{\sqrt{\alpha}}{0.135 + \sqrt{0.98 + \alpha}}$\cite{Tcorrection2}, which denotes the fraction of the latent heat transformed into the bulk motion of the fluid. Besides, efficiency $\kappa_v$ will be determined by the bubble expansion mode.
\subsubsection{Turbulence}
Since the plasma is ionized fully, percolation can also induce MHD turbulence in the plasma.
Then, the GW contribution from the turbulence can be written as\cite{GWtrub1,GWtrub2}
\begin{eqnarray}
&&h^2\Omega_{turb}(f)= 3.35\times10^{-4}v_b\Big(\frac{\beta}{H_n}\Big)^{-1}
\Big(\frac{\kappa_{turb}\alpha}{1+\alpha}\Big)^{3/2}\Big(\frac{g_*}{100}\Big)^{-1/3}\nonumber\\
&&\hspace{2cm}\times\frac{\Big(f/{f_{turb}}\Big)^3}{\Big({1+ f/f_{turb}}\Big)^{11/3}\Big({1+ {8\pi f}/{h_{n}}}\Big)},
\label{GWtrubf}
\end{eqnarray}
where $h_n = 1.65\times10^{-5}\Big(\frac{T_n}{100{\rm GeV}}\Big)\Big(\frac{g_*}{100}\Big)^{1/6}\;{\rm Hz}$ is the Hubble parameter today. We set the efficiency factor of the latent heat for turbulence to be $\kappa_{turb}\simeq 0.05\kappa_v$\cite{GWgeneration2}. The peak frequency $f_{turb}$ of the GWs produced from the turbulence will be as follows
\begin{eqnarray}
f_{turb}=2.7\times10^{-5}\frac{1}{v_b}\Big(\frac{\beta}{H_n}\Big)
\Big(\frac{T_n}{100\;{\rm GeV}}\Big)\Big(\frac{g_*}{100}\Big)^{1/6}\;{\rm Hz}.
\label{GWtrubftrub}
\end{eqnarray}
\section{the numerical results of GW spectrum in the B-LSSM}
In our numerical calculation, we consider the constraints of parameter space related to the future experiment within B-LSSM. In section II, we study the Higgs masses with the one-loop zero temperature effective potential corrections. Parameters $g_B,~g_{YB},~tb$ and $tb'$ have been limited within small regions, which will affect the numerical results of GW spectrum. As well as, we have studied Higgs decay modes, $B$ meson rare decay $\bar{B}\rightarrow X_s\gamma$ and the muon anomalous magnetic dipole moment in the B-LSSM\cite{g-22021}. The corresponding parameter constraints will be considered in our numerical discussion. In the following numerical discussion, we use the numerical package \textbf{CosmoTransitions}\cite{CosmoTransitions} for analyzing the corresponding PT. After calculating, we obtain the bounce solutions and discover the nucleation temperature $T_n$. The value of $S_E/{T_n}$ is around 140 below certain nucleation temperature $T_n$. Then we can obtain the important results of parameters $\alpha$ and $\beta$, which finally determine the GW spectrums.

\begin{figure}[t]
\centering
\includegraphics[width=8cm]{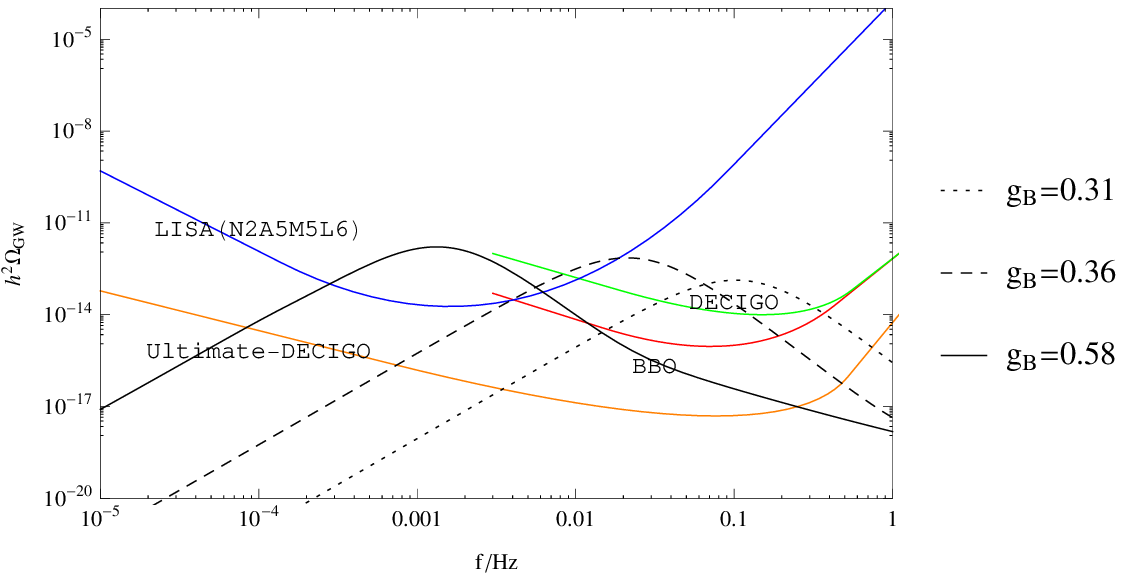}
\includegraphics[width=8cm]{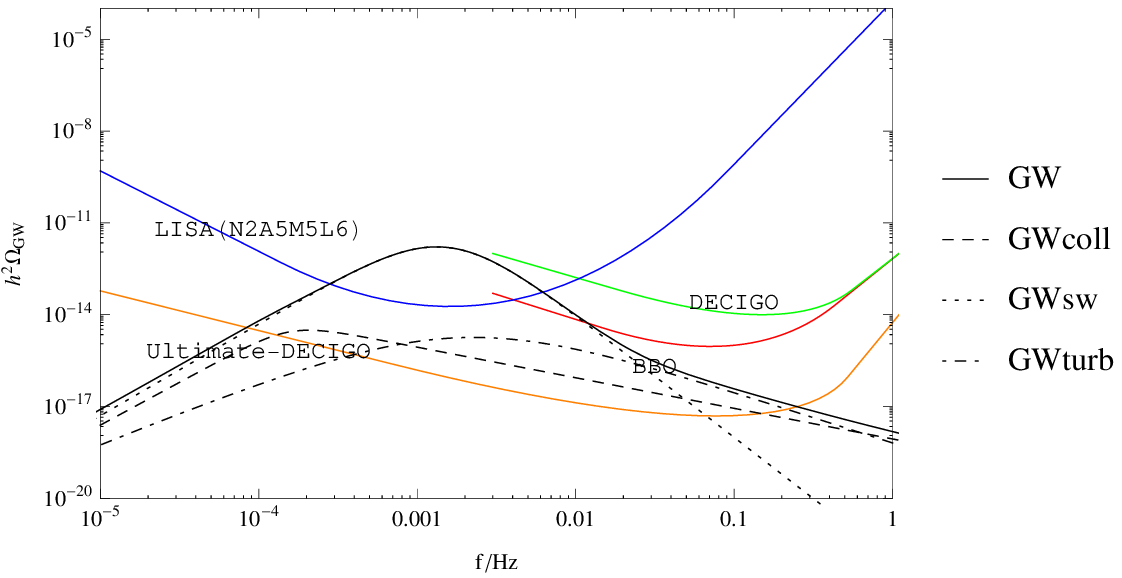}
\caption{The contributions of GW spectrum vary with parameters $g_B$ in the B-LSSM.}
 \label{gB}
\end{figure}

In this part, we study the dependence of the GW spectrums on the parameters $tb$, $g_{YB}$, $g_B$, $\mu_{\eta}$, $B_{\eta}$, $m_{\eta}$ and $m_{\bar{\eta}}$. First, the GW spectrums versus parameter $g_B$ are plotted in FIG.\ref{gB} with $tb=10$, $g_{YB}=-0.1$, $\mu_{\eta}=380~\rm{GeV}$, $B_{\eta}=8\times10^5~\rm{GeV^2}$, $m_{\eta}=1500~\rm{GeV}$ and $m_{\bar{\eta}}=300~\rm{GeV}$. The color lines show the sensitivities of the future GW experiments, which include the LISA(N2A5M5L6), BBO, DECIGO and Ultimate-DECIGO. We can discover that the present and future GW observations could probe a broad parameter space for $g_B$. Besides, the GWs signal can be detected by the LISA, BBO, DECIGO and Ultimate-DECIGO with $g_B$ taking the value around 0.36. Then, the GW spectrums $h^2\Omega_{GW}$ versus frequency $f$ will be further researched in the right of FIG.\ref{gB} with $g_B=0.58$, which is preferred by the space-based experiments. The contributions to bubble collision, sound waves and turbulence are respectively plotted by dashed, dotted and dot dashed lines. The contributions of GW spectrum, coming from the sum of the aforementioned three contributions, will be denoted by the solid black line. We can easily find that the contribution to sound wave is dominant and almost indistinguishable from the solid black line when $f$ is around $10^{-4}\sim0.01~{\rm Hz}$.

\begin{figure}[t]
\centering
\includegraphics[width=8cm]{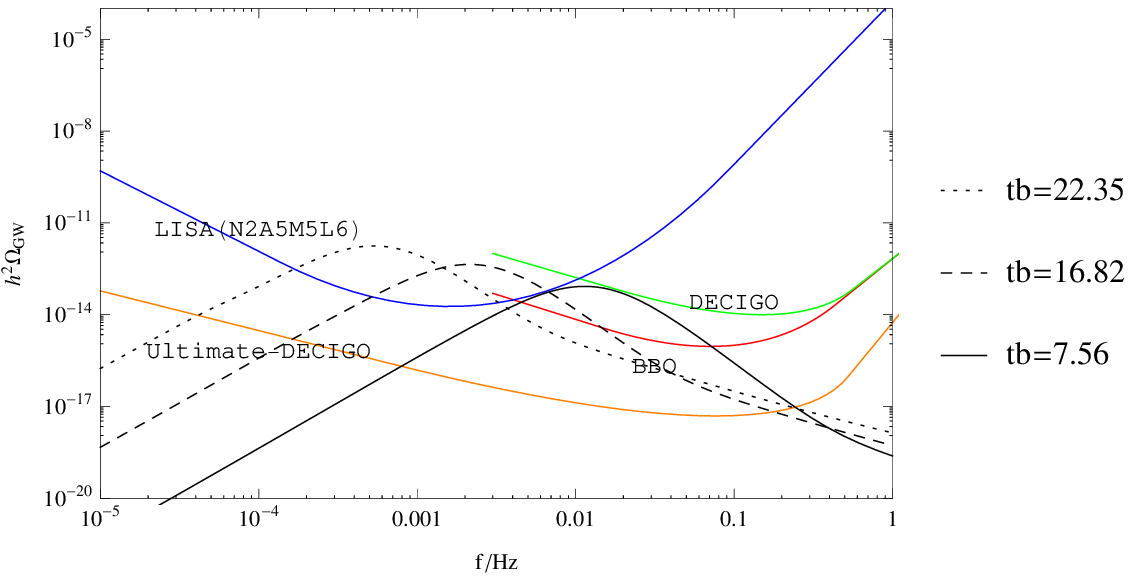}
\includegraphics[width=8.2cm]{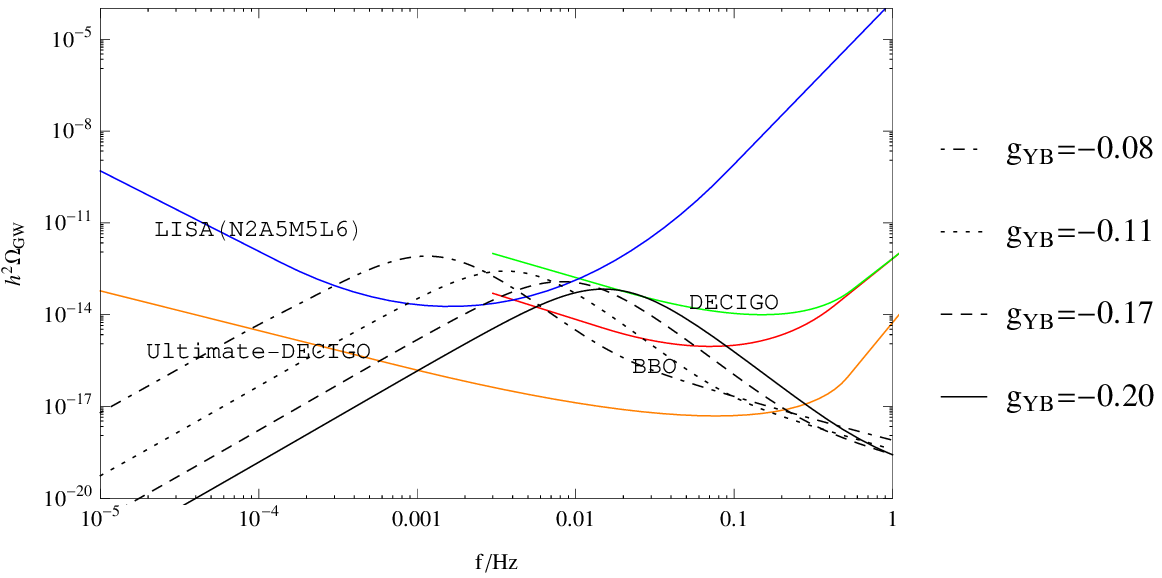}
\caption{The contributions of GW spectrum vary with parameters $tb$ and $g_{YB}$ in the B-LSSM.}
 \label{tbgYB}
\end{figure}

Then, the GW spectrums versus parameters $tb$ and $g_{YB}$ are plotted in FIG.\ref{tbgYB} with $g_B=0.6$, $\mu_{\eta}=380~\rm{GeV}$, $B_{\eta}=8\times10^5~\rm{GeV^2}$, $m_{\eta}=1500~\rm{GeV}$ and $m_{\bar{\eta}}=300~\rm{GeV}$. Parameter $tb$ presents in the Yukawa couplings $Y_t$ and $Y_b$, which affects the the finite temperature effective potential through the temperature corrections of the Higgs boson and third squark masses, so that influence the GWs. We can discover that the GW spectrums can be detected by the LISA, BBO and Ultimate-DECIGO as $8<tb<23$ and $-0.17<g_{YB}<-0.08$. In our following discussion, we make $tb=10$ and $g_{YB}=-0.1$.

\begin{figure}[t]
\centering
\includegraphics[width=8cm]{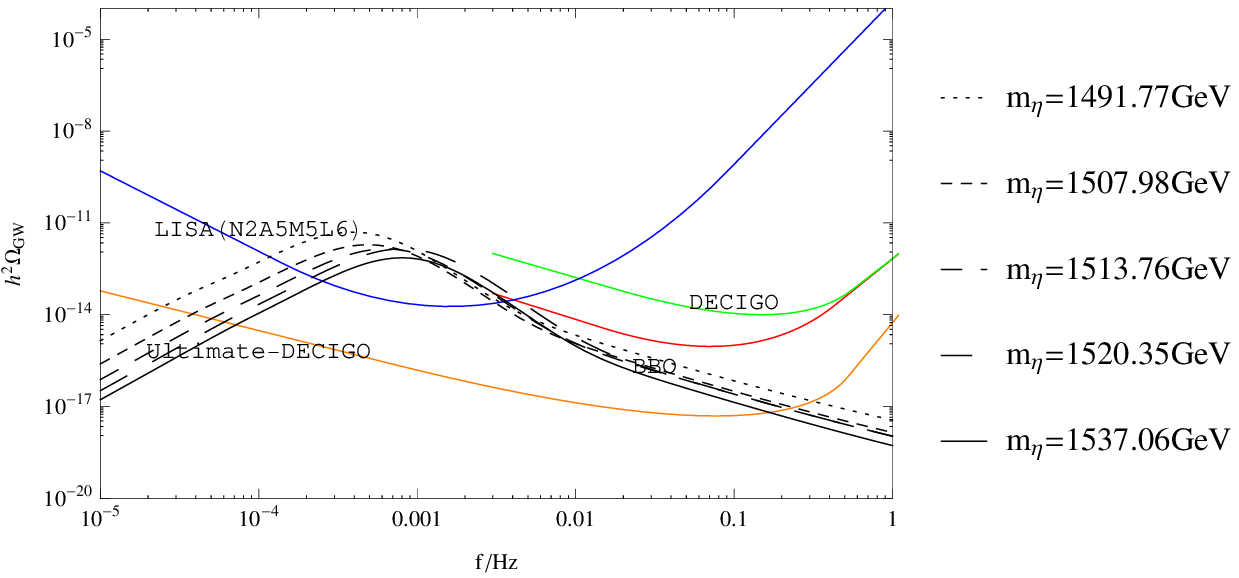}
\includegraphics[width=8cm]{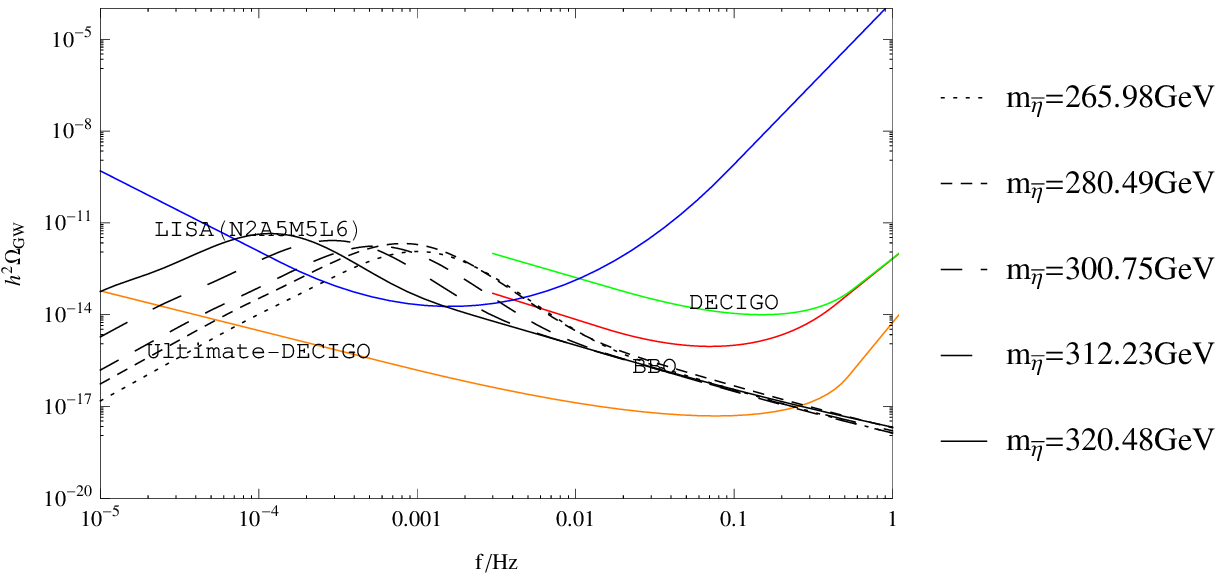}
\caption{The contributions of GW spectrum vary with parameters $m_{\eta}$ and $m_{\bar{\eta}}$ in the B-LSSM.} \label{mytaytabar}
\end{figure}

Parameters $m_{\eta}$ and $m_{\bar{\eta}}$ both exist in the tree-level zero temperature effective potential. When $tb=10$, $g_{YB}=-0.1$, $g_B=0.6$, $\mu_{\eta}=380~\rm{GeV}$ and $B_{\eta}=8\times10^5~\rm{GeV^2}$, the GW spectrums versus frequency $f$ with parameters $m_{\eta}$ and $m_{\bar{\eta}}$ will be shown in FIG.\ref{mytaytabar}. The smaller $m_{\eta}$ but larger $m_{\bar{\eta}}$, the larger GW spectrum with relatively smaller peak frequency we can obtain. Other than this, the value of parameter $m_{\eta}$ is around $1500~\rm{GeV}$, while the value of parameter $m_{\bar{\eta}}$ is around $300~\rm{GeV}$. Our numerical simulations reveal that the parameters $m_{\eta}$ and $m_{\bar{\eta}}$ possess tiny parameter spaces to acquire suitable GW spectrums.

\begin{figure}[t]
\centering
\includegraphics[width=8cm]{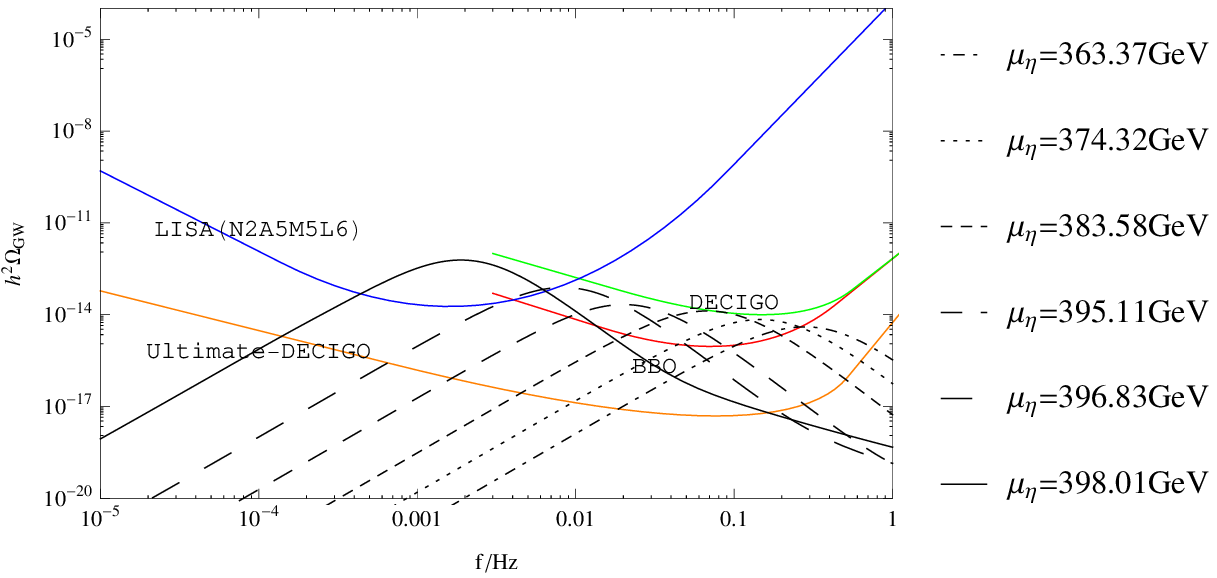}
\includegraphics[width=8.1cm]{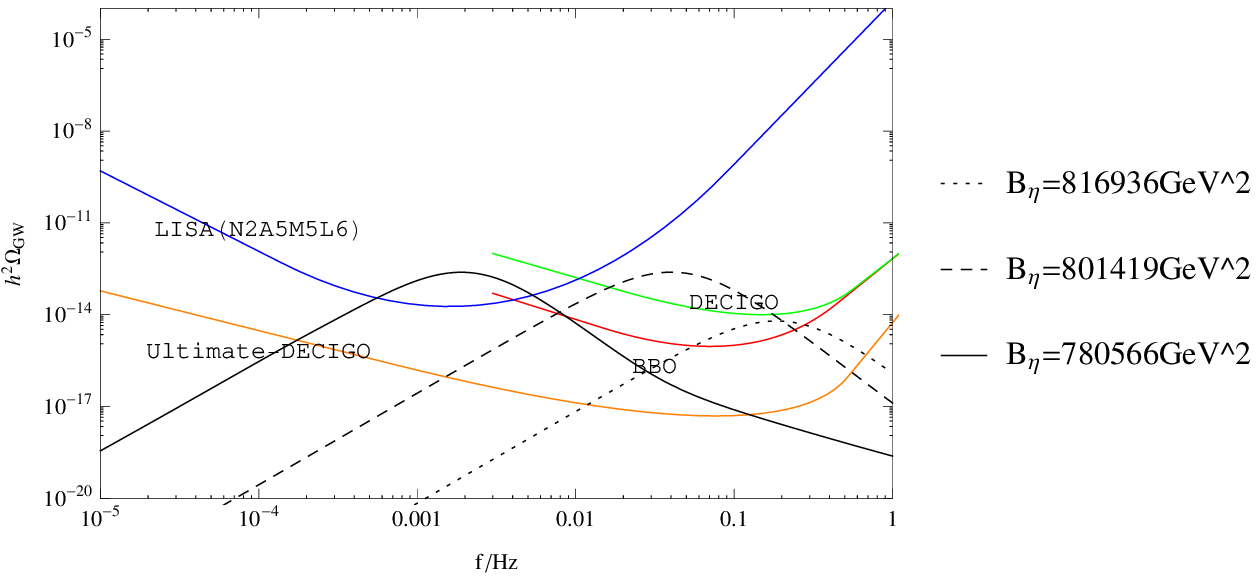}
\caption{The contributions of GW spectrum vary with parameters $\mu_{\eta}$ and $B_{\eta}$ in the B-LSSM.} \label{muBBmuB}
\end{figure}

After fixing $tb=10$, $g_{YB}=-0.1$, $g_B=0.6$, $m_{\eta}=1500~\rm{GeV}$ and $m_{\bar{\eta}}=300~\rm{GeV}$ in the B-LSSM, the GW spectrums versus parameters $\mu_{\eta}$ and $B_{\eta}$ are demonstrated through the FIG.\ref{muBBmuB}. As $\mu_{\eta}$ is in the region of $360\sim400~\rm{GeV}$, a larger value of $\mu_{\eta}$ tends to produce a larger GW spectrum with a relatively smaller peak frequency. It is worth noting that GW spectrum can be detected by the LISA, BBO and Ultimate-DECIGO with $\mu_{\eta}\simeq400~\rm{GeV}$. Then the GW spectrum versus parameter $B_\eta$ will be studied. With the increasing value of $B_\eta$, the GW signal gradually strengthens. As $B_\eta\simeq8\times10^5~\rm GeV^2$, GW spectrum will be detected by DECIGO, BBO and Ultimate-DECIGO; While $B_\eta\simeq7.8\times10^5~\rm GeV^2$, GW signal will be detected by LISA, BBO and Ultimate-DECIGO.

\begin{figure}[t]
\centering
\includegraphics[width=8cm]{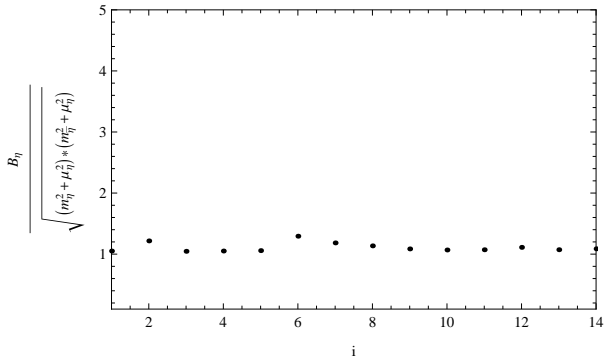}
\caption{The ratio of the random parameter $B_{\eta}$ to $\sqrt{(m_{\eta}^2+\mu_{\eta}^2)(m_{\bar{\eta}}^2+\mu_{\eta}^2)}$.} \label{finetune}
\end{figure}

Through the above discussion in FIG.\ref{mytaytabar} and FIG.\ref{muBBmuB}, we find that parameters $m_{\eta}$, $m_{\bar{\eta}}$, $\mu_{\eta}$ and $B_{\eta}$ fluctuate within a small and fine-tuned region. We expect to find the relationship between these parameters to explain these relatively special parameter values. So, we random scan the parameter space as $7<tb<25$, $-0.23<g_{YB}<-0.07$, $0.3<g_B<0.7$, $200~\rm{GeV}<\mu_{\eta}<500~\rm{GeV}$, $7\times10^5~\rm{GeV^2}<B_{\eta}<10^6~\rm{GeV^2}$, $1400~\rm{GeV}<m_{\eta}<2000~\rm{GeV}$ and $100~\rm{GeV}<m_{\bar{\eta}}<400~\rm{GeV}$. When the numerical results satisfy the strong first-order PT and can obtain the suitable GW signals, we plot the ratio of random parameter $B_{\eta}$ to $\sqrt{(m_{\eta}^2+\mu_{\eta}^2)(m_{\bar{\eta}}^2+\mu_{\eta}^2)}$ in FIG.\ref{finetune}. We conclude that these parameters satisfy the approximate relation: $\frac{B_{\eta}}{\sqrt{(m_{\eta}^2+\mu_{\eta}^2)(m_{\bar{\eta}}^2+\mu_{\eta}^2)}}\simeq1.1$.

\section{conclusion}
In the B-LSSM, we first study the Higgs masses with the one-loop zero temperature effective potential corrections. When the physical Higgs mass satisfies $3\sigma$ experimental interval, parameters $g_B,~g_{YB}$ and $tb$ are limited within small regions: $0.3\leq g_B \leq0.7$, $-0.23\leq g_{YB} \leq-0.07$ and $7\leq tb\leq25$. In additional, the larger parameter $B_{\eta}$, the larger second-light Higgs mass we can obtain.
Then, we study the GW spectrums generated by the strong first-order PT with the two $U(1)_{B-L}$ singlet Higgs superfields. Our numerical calculations reveal that the contribution from sound wave is dominated to generate the GW signals when the frequency $f$ is around $10^{-4}\sim0.01~{\rm Hz}$. Parameters $g_B$ and $tb$ affect the GW signals obviously within the regions: $0.3<g_B<0.6$ and $8<tb<23$. Besides, we can easily find that parameters $m_{\eta}$, $m_{\bar{\eta}}$, $\mu_{\eta}$ and $B_{\eta}$ exist relationship: $\frac{B_{\eta}}{\sqrt{(m_{\eta}^2+\mu_{\eta}^2)(m_{\bar{\eta}}^2+\mu_{\eta}^2)}}\simeq1.1$. As the parameters take the suitable values, the strength of the GW spectrum will be as large as $h^2\Omega_{GW}\sim10^{-11}$, which may be detected by the future GW detection experiments such as LISA, BBO, DECIGO and Ultimate-DECIGO.

{\bf Acknowledgments}

We are very grateful to Li-Gong Bian the teacher of Chongqing University and Wu-Long Xu the Dr. of Beijing University of Technology, for giving us some useful discussions. This work is supported by the Major Project of National Natural Science Foundation of China (NNSFC) No. 12075074, No. 11535002 and No. 11705045, Natural Science Foundation of Hebei Province No. A2020201002 and the youth top-notch talent support program of the Hebei Province.

\appendix
\section{}
B-LSSM is assumed the gauged one, and we have proved carefully that this model is anomaly free. We use $Y^Y$ and $Y^{B-L}$ respectively representing the charge of $U(1)_Y$ and $U(1)_{B-L}$. The quantum field theory written by M. E. Peskin has introduced the anomaly free of standard model(SM) in page 705-707. So, we show the demonstration of anomaly free in the B-LSSM.

1. In the B-LSSM, the anomaly of three $SU(2)$ bosons banishes, as well as the anomaly of three $SU(3)$ boson. These anomalies are same as the SM.

2. In the B-LSSM, the anomalies containing one $SU(3)$ boson or one $SU(2)$ boson are proportional to $Tr[t^a]=0$ or $Tr[\tau^a]=0$.

3. The remaining nontrivial anomalies are: the anomaly of one $U(1)$ boson with two $SU(3)$ bosons, the anomaly of one $U(1)$ boson with two $SU(2)$, the anomaly of three $U(1)$ bosons and the gravitational anomaly with one $U(1)$ boson. In the B-LSSM, there are $U(1)_Y$ and $U(1)_{B-L}$ gauge groups, which are more complicated than the SM.

(1) The anomaly of one $U(1)_Y$ or $U(1)_{B-L}$ boson with two $SU(3)$ bosons is proportional to the group theory factor $Tr[t^at^bY^Y]=\frac{1}{2}\delta^{ab}\sum_qY_q^Y$ or $Tr[t^at^bY^{B-L}]=\frac{1}{2}\delta^{ab}\sum_qY_q^{B-L}$.

(2) Similarly, the anomaly of one $U(1)_Y$ or $U(1)_{B-L}$ boson with two $SU(2)$ bosons is also proportional to the group theory factor $Tr[\tau^a\tau^bY^Y]=\frac{1}{2}\delta^{ab}\sum_{fL}Y_{fL}^Y$ or $Tr[\tau^a\tau^bY^{B-L}]=\frac{1}{2}\delta^{ab}\sum_{fL}Y_{fL}^{B-L}$.

(3) The anomalies of three $U(1)$ gauge bosons are divided into four types with the B-LSSM: $Tr[Y^YY^YY^Y]=\sum_n(Y_n^Y)^3$, $Tr[Y^{B-L}Y^{B-L}Y^{B-L}]=\sum_n(Y_n^{B-L})^3$, $Tr[Y^YY^{B-L}Y^{B-L}]=\sum_n(Y_n^Y)(Y_n^{B-L})^2$ and $Tr[Y^YY^YY^{B-L}]=\sum_n(Y_n^Y)^2(Y_n^{B-L})$.

(4) the gravitational anomaly with one $U(1)$ gauge boson $U(1)_Y$ or $U(1)_{B-L}$ is proportional to $Tr[Y^Y]=\sum_n(Y_n^Y)$ or $Tr[Y^{B-L}]=\sum_n(Y_n^{B-L})$.

The anomalies that do not relate with $U(1)_{B-L}$ are proved free and they are very similar as the SM condition.
\begin{eqnarray}
&&\sum_qY_q^Y=[\frac{1}{6}\times2+\frac{1}{3}+(-\frac{2}{3})]\times3\times3=0,\nonumber\\
&&\sum_{fL}Y_{fL}^Y=\frac{1}{6}\times2\times3\times3+(-\frac{1}{2})\times2\times3=0,\nonumber\\
&&\sum_n(Y_n^Y)^3=[(\frac{1}{6})^3\times2+(\frac{1}{3})^3+(-\frac{2}{3})^3]\times3\times3+[(-\frac{1}{2})^3\times2+1^3]\times3=0,\nonumber\\
&&\sum_nY_n^Y=[\frac{1}{6}\times2+\frac{1}{3}+(-\frac{2}{3})]\times3\times3+[(-\frac{1}{2})\times2+1]\times3=0.\nonumber
\end{eqnarray}
The anomalies including $U(1)_{B-L}$ are also proved free, which are more complicated than the calculation of SM.
\begin{eqnarray}
&&\sum_qY_q^{B-L}=[\frac{1}{6}\times2+(-\frac{1}{6})+(-\frac{1}{6})]\times3\times3=0,\nonumber\\
&&\sum_{fL}Y_{fL}^{B-L}=\frac{1}{6}\times2\times3\times3+(-\frac{1}{2})\times2\times3=0,\nonumber\\
&&\sum_n(Y_n^{B-L})^3=[(\frac{1}{6})^3\times2+(-\frac{1}{6})^3+(-\frac{1}{6})^3]\times3\times3+[(-\frac{1}{2})^3\times2+(\frac{1}{2})^3+(\frac{1}{2})^3]\times3=0,\nonumber\\
&&\sum_nY_n^Y(Y_n^{B-L})^2=[\frac{1}{6}\times(\frac{1}{6})^2\times2+\frac{1}{3}\times(-\frac{1}{6})^2+(-\frac{2}{3})^2\times(-\frac{1}{6})]\times3\times3\nonumber\\
&&\hspace{2.8cm}+[(-\frac{1}{2})^2\times(-\frac{1}{2})\times2+1\times(\frac{1}{2})^2]\times3=0,\nonumber\\
&&\sum_n(Y_n^Y)^2(Y_n^{B-L})=[(\frac{1}{6})^2\times\frac{1}{6}\times2+(\frac{1}{3})^2\times(-\frac{1}{6})+(-\frac{2}{3})\times(-\frac{1}{6})^2]\times3\times3\nonumber\\
&&\hspace{3.1cm}+[(-\frac{1}{2})\times(-\frac{1}{2})^2\times2+1^2\times\frac{1}{2}]\times3=0,\nonumber\\
&&\sum_nY_n^{B-L}=[\frac{1}{6}\times2+(-\frac{1}{6})+(-\frac{1}{6})]\times3\times3+[(-\frac{1}{2})\times2+\frac{1}{2}+\frac{1}{2}]\times3=0.\nonumber
\end{eqnarray}

 \end{document}